# Tubulin bond energies and microtubule biomechanics determined from nanoindentation *in silico*

Olga Kononova,[†,‡] Yaroslav Kholodov,[‡] Kelly E. Theisen,[§,] Kenneth A. Marx,[†] Ruxandra I. Dima,[§] Fazly I. Ataullakhanov,[|,⊥,] Ekaterina L. Grishchuk,[○,*], and Valeri Barsegov[†,‡,*]

[†]Department of Chemistry, University of Massachusetts, Lowell, MA 01854, USA; [‡]Moscow Institute of Physics and Technology, Moscow region, 141700, Russia; [§]Department of Chemistry, University of Cincinnati, Cincinnati, OH 45221, USA; [|]Center for Theoretical Problems of Physicochemical Pharmacology, Russian Academy of Sciences, Moscow 119991, Russia; [⊥]Physics Department, Moscow State University, Moscow 119991, Russia; [○]Physiology Department, Perelman School of Medicine, University of Pennsylvania, Philadelphia, PA 19104

[*]Corresponding authors: Valeri_Barsegov@uml.edu, gekate@mail.med.upenn.edu

# ABSTRACT

Microtubules, the primary components of the chromosome segregation machinery, are stabilized by longitudinal and lateral non-covalent bonds between the tubulin subunits. However, the thermodynamics of these bonds and the microtubule physico-chemical properties are poorly understood. Here, we explore the biomechanics of microtubule polymers using multiscale computational modeling and nanoindentations *in silico* of a contiguous microtubule fragment. A close match between the simulated and experimental force-deformation spectra enabled us to correlate the microtubule biomechanics with dynamic structural transitions at the nanoscale. Our mechanical testing revealed that the compressed MT behaves as a system of rigid elements interconnected through a network of lateral and longitudinal elastic bonds. The initial regime of continuous elastic deformation of the microtubule is followed by the transition regime, during which the microtubule lattice undergoes discrete structural changes, which include first the reversible dissociation of lateral bonds followed by irreversible dissociation of the longitudinal bonds. We have determined the free energies of dissociation of the lateral (6.9±0.4 kcal/mol) and longitudinal (14.9±1.5 kcal/mol) tubulin-tubulin bonds. These values in conjunction with the large flexural rigidity of tubulin protofilaments obtained (18,000- 26,000 pN·nm$^2$), support the idea that the disassembling microtubule is capable of generating a large mechanical force to move chromosomes during cell division. Our computational modeling offers a comprehensive quantitative platform to link molecular tubulin characteristics with the physiological behavior of microtubules. The developed *in silico* nanoindentation method provides a powerful tool for the exploration of biomechanical properties of other cytoskeletal and multiprotein assemblies.

# INTRODUCTION

Microtubules (MTs) are essential for health and viability of eukaryotic cells. Stable MTs are fairly rigid,[1] which enables them to serve as important structural and organizing elements. MTs form long and durable linear tracks for neuronal transport, and the mechanical properties of MTs help to define cell architecture and polarity.[2] The dynamics of MTs, i.e. their ability to undergo stochastic cycles of polymerization and depolymerization, also play a prominent role in many cellular processes.[3,4] MTs play a vital role during cell division, when they form a mitotic spindle;[5] as a result, different MT disrupting or stabilizing drugs are widely used as chemotherapeutic agents.[6] Importantly, the disassembling MTs have been proposed to serve as a primary biological motor for poleward chromosome motion during mitosis.[7,8] However, understanding the underlying mechanisms for different MT functions is impeded by a lack of quantitative knowledge about the thermodynamics and biomechanics of these complex cytoskeletal structures.

MTs are hollow protein cylinders that contain lateral assemblies of protofilaments: the linear strands of longitudinally-arranged αβ-tubulin dimers (Fig 1A).[9] A biologically relevant form of MT contains 13 protofilaments that are arranged in a left-hand 3 start helix. Such a multi-protofilament structure makes it difficult to establish a direct correspondence between molecular tubulin characteristics and observed MT properties *in vitro*. Theoretical approaches have played an important role in providing such a link, for example, by exploring different mechanisms of MT dynamics and force generation.[10-13] Theoretical studies have identified several microscopic properties that are critically important for understanding the MT behavior. These features include the thermodynamic characteristics of interfaces between adjacent tubulins, i.e. the energy of lateral and longitudinal bonds, and the mechanical flexural rigidity of individual tubulin protofilaments. To date, it has not been possible to probe these properties via direct experimental measurements, which explains why virtually every aspect of MT thermodynamics and mechanics is still controversial.

It is generally agreed that in the MT lattice the longitudinal tubulin-tubulin bonds are stronger than the lateral bonds, because during MT disassembly the lateral tubulin bonds dissociate prior to the longitudinal ones. This is evident from the presence of curled "ram horns" at the ends of shortening polymers[14] and is in agreement with computer calculations based on tubulin structure.[15] However, the absolute values of the tubulin-tubulin bond energies are not known. Traditional thermodynamic analyses of these bonds have led to ambiguous results mainly because of the complexity of pathways for tubulin assembly and disassembly.[16] Indeed, the binding rate constants for tubulin attachment to various sites at the ragged MT tip can vary due to the differences in the number and location of neighboring subunits.[17] Additional difficulty concerns the dissociation rate, which can be affected by the number of lateral contacts for a given dimer, as well as by the rigidity of MT protofilaments, which bend concomitantly with tubulin disassembly. The energy of lateral and longitudinal bonds' dissociations have previously been estimated using different kinetic and mechanical MT models; the energy values vary significantly from 3 to 15 $k_BT$ for the lateral bonds, and from 6 to 20 $k_BT$ for the longitudinal bonds.[18-23] Quantum calculations have also been employed but the obtained estimates are unrealistically large (up to 186 $k_BT$ for the lateral bonds and 158 $k_BT$ for the longitudinal bonds[24,25]). The shapes of the free energy profiles and even the geometry and number of the sites for tubulin-tubulin interactions in the MT models are debated.[18,20,22,23,26]

The flexural rigidity of MT protofilaments is also a subject of debate. Previous theoretical estimates of this quantity vary by an order of magnitude, from 1,500 to 28,000 pN nm$^2$,[27,28] which correspond to energies of 3.7 to 64 $k_BT$ per dimer for full protofilament straightening. Accurate determination of protofilament rigidity is experimentally difficult because protofilaments are fragile transient structures. Knowing flexural rigidity, however, is important, because it has direct implications for mechanisms of force generation during MT depolymerization. Indeed, MT depolymerization can generate a large force *in vitro* and *in vivo*,[29,30] but the underlying mechanism is controversial[27]. In the power stroke-based mechanism of force generation, the bending protofilaments are thought to transmit a large force available from tubulin-tubulin energetics, but this energy can be used to move the associated cargo only if protofilaments are fairly rigid.[10]

The various functional roles played by MTs in eukaryotic cells necessitate rigorous quantitative analysis of their thermodynamic and mechanical properties. Recently, considerable progress has been achieved in the mechanical testing of biological protein assemblies,[31] including the MT response to compressive force.[32,33] Such dynamic force measurements present a unique methodology to deform or rupture the non-covalent bonds of the MT lattice, opening an experimental avenue to determine the underlying microscopic characteristics. The published experimental force-indentation spectra for the MT reveal a complex multi-step deformation mechanism. [32,33] However, a detailed structure-based interpretation of these results has been lacking. Here, we have carried out the controlled *in silico* nanoindentations of the MT by combining Molecular Dynamics (MD) simulations accelerated on Graphics Processing Units (GPUs)[34,35] of the atomic tubulin structure and the $C_\alpha$-based Self Organized Polymer (SOP) model[36-40] of the MT fragment, which contains 13 protofilaments, each 8 tubulin dimers in length (Fig. 1). The computational acceleration on Graphics Processing Units (GPUs) has enabled us to apply the experimentally relevant force-loading rate (cantilever velocity $v_f$ = 1.0 µm/s) and to span the experimental timescale (~50 ms). Close agreement between experimental and simulated force spectra has allowed us to resolve structural transitions in the MT lattice that underpin the MT lattice biomechanics in the experimentally inaccessible sub-nanometer scale of length. Importantly, using our novel methodology of nanoindentation *in silico* we were able to directly calculate the energies of lateral and longitudinal tubulin-tubulin contacts and to obtain an independent estimate of the flexural rigidity of single tubulin protofilaments.

**RESULTS**

**SOP model provides accurate description of the experimental force-indentation spectra:** The simulated force-indentation spectra, i.e. the profiles of the indentation force *F* vs. the cantilever tip displacement (indentation depth) *X* (the *FX* curves) and the profiles of *F* vs. the virtual cantilever base (or piezo) displacement *Z* (the *FZ* curves) are presented in Fig. 2A. Importantly, these curves are very similar to the corresponding experimental spectra (see Fig. 1C in Ref. (32)). The *FZ* and *FX* curves exhibit the single-step transitions, characterized by a single force peak, and multi-step transitions with several force peaks. Although the force spectra show some variability depending on the location of indentation points, each spectrum reveals three distinct regimes (see Supporting Movie S1 is Supporting Information (SI)): (1) the linear-like regime of continuous elastic deformation ($Z < 15$-$20$ nm; $X < 6$-$8$ nm); (2) the transition regime where the MT lattice undergoes discrete structural transitions ($15$-$20$ nm $< Z < 25$-$30$ nm; $6$-$8$ nm $< X < 11$-$13$ nm); and (3) the post-collapse regime ($Z > 25$-$30$ nm; $X > 11$-$13$ nm) (Fig. 2). We estimated the spring constant $K_{MT}$ from the initial slope of the $FZ$ curves (linear-like regime), and extracted the values of critical force $F^*$ (peak force in $FZ$ curves) and the critical distance $Z^*$, at which the transition to the collapsed state occurs. These values are in good quantitative agreement with their experimental counterparts (compared in Table 1). The slightly higher theoretical values of $F^*$ and $Z^*$ are due to our using a faster cantilever velocity ($v_f = 1.0$ µm/s vs. $v_f \approx 0.2$ µm/s used in Ref. (32,33)). Thus, the SOP model provides a very good description of the physicochemical properties of the MT lattice. In the simulations described above, we imposed hard constraints at the ends of the MT fragment (see Materials and Methods) to mimic the long persistence length of the MT polymers (microns to millimeters). However, the force spectra were very similar when the soft (harmonic) constraints were applied, and the relative difference between the values of $F^*$, $Z^*$ and $X^*$ from simulations with soft constraints and hard constraints were within the standard deviations (data not shown).

**MT is a network of rigid elements interconnected via elastic lateral and longitudinal bonds:** Simulations for 7 indentation points were carried out using tips of different size. The summarized description of all observed transitions is provided in Table S1. Comparison of the $FZ$ and $FX$ curves for a 10 nm tip (Fig. 2) vs. a 15 nm tip (Fig. S1) shows that the force spectra are similar, although the values of $F^*$, $X^*$ (critical indentation depth), and $K_{MT}$ increase slightly with tip size (Table 1). Consider examples of the forward indentation for the 10 nm tip followed by tip retraction at the surface of a protofilament (indentation points 2 and 3; Figs. 3A-3B) and between protofilaments (indentation points 6 and 7; Figs. 3C-3D). The critical force $F^*$ and critical indentation depth $X^*$ depend on where the compressive force is applied: $F^* = 0.65$-$0.7$ nN and $X^* \approx 12$ nm for compressing a protofilament (Fig. 3A) are larger than $F^* = 0.5$-$0.55$ nN and $X^* \approx 10$ nm for compressing the interface between protofilaments (Fig. 3C). We also profiled the slope of the $FX$ curve ($dF/dX$) – a measure of mechanical compliance of the MT. We found that $dF/dX$ varies largely with $X$ (Figs. 3B and 3D): steep increases interrupted by sudden drops of $dF/dX$ indicate that the MT lattice behaves as a soft material. The heights of the peaks of $dF/dX$ mark the limits of deformability of the MT cylinder. The MT resists the mechanical collapse longer (strong last peak of *dF/dX*) when indented on the protofilament rather than between the protofilaments (weak last peak of *dF/dX*). This indicates that the lateral interfaces between tubulins are softer (more compliant mechanically) than between longitudinal tubulins within a protofilament. We arrive at similar conclusions when considering the results obtained with a 15 nm tip for indentation points 1, 3, and 5, 7 (Fig. S2). The profiles of structure overlap $\chi$ (defined in SI) show that the collapsed MT lattice remains ~80-90% similar to the uncompressed state (*the insets* to Figs. 3B, 3D, and S2B, S2D). This implies that stress-dependent changes are mainly localized to the lateral and longitudinal interfaces. Hence, the network of lateral and longitudinal bonds is the origin of elasticity for the MT lattice.

**MT deformation and collapse occur via specific multi-step mechanism:** Our study demonstrates a qualitative similarity between the *FX* curves and profiles of *dF/dX* and $\chi$ for different indentation points (Figs. 2, 3 for 10 nm tip, and Figs. S1, S2 for 15 nm tip). Analysis of structures generated under different indentation conditions has revealed that the MT transition to the collapsed state occurs by a surprisingly conserved pathway (Supporting Movie S1), which we illustrate for two examples of MT indentation (Fig. 4). Initially, the MT lattice resists deformation, as seen from the increase of *dF/dX* (*the inset* in Fig. 4), which results in small variations in the local curvature of the MT cylinder under the tip (Fig. S3A). This is the linear-like regime of (continuous) elastic deformation as evidenced from the quasi-linear dependence of *F* on *X* (white region in Fig. 4); this regime persists until $X \approx 6\text{-}8$ nm (structure 1 in Figs. 2, S1, and 4; see Fig. S3A). The compressive force loads an increasingly larger portion of the MT surface leading to the MT cylinder flattening ("buckling"). Indentation beyond $X \approx 6\text{-}8$ nm can no longer be accommodated by the MT bending alone. At this point, the MT system enters the transition regime (gray region in Fig. 4) in which discrete structural changes occur. In this regime, mechanical tension exceeds the strength of lateral and longitudinal bonds, which results in their sequential rupture: the lateral bonds dissociate first at $X \approx 6\text{-}8$ nm (structure 2a, Fig. 4), and the longitudinal bonds dissociate second at $X \approx 11\text{-}13$ nm (structure 2b, Fig. 4). The latter event triggers the MT lattice rapid transitioning to the collapsed state, which results in a sharp force drop (force peak in Fig. 4). This crossover from the continuous deformation to the multi-step discrete dissociation transitions was observed in all indentation simulations (21 runs), regardless of where the compressive force was applied. Importantly, we detected the dissociation of the longitudinal inter-dimer bonds but not the intra-dimer bonds, consistent with tubulin heterodimer being a major structural unit for MT disassembly. Disruption of the lateral interfaces between the α-tubulins occurred simultaneously with loss of lateral contacts between the β-tubulins. Beyond $X \approx 20$ nm indentation, which corresponds to the post-collapse regime, the tip indented the lower portion of the MT cylinder (not shown), and the resulting events were not analyzed.

**Dissociation of the lateral but not longitudinal contacts is reversible:** We carried out simulations of the force-quenched tip retraction with 10 nm tip (Figs. 2, 3) and 15 nm tip (Figs. S1, S2), in which we reversed the direction of cantilever motion thereby gradually decreasing to zero the amplitude of compressive force (see Supporting Movie S2). We used the MT structures from the simulations of forward deformation for $X$ = 7, 11 and 21 nm indentation. These structures correspond to the buckled MT ($X \approx 7$ nm), the MT with disrupted lateral bonds ($X \approx 11$ nm), and the MT with disrupted lateral and longitudinal bonds ($X \approx 21$ nm). To monitor the progress of MT lattice remodeling, we analyzed the structure overlap $\chi$ (*the inset* to Figs. 2, S1). In full agreement with experiment (Figs. 1, 5 in Ref. (33)), we found that the deformation is fully reversible for small indentations ($X < 7$ nm), partially reversible with small hysteresis for larger indentations (7 nm $< X <$ 11 nm), and irreversible with large hysteresis for indentations larger than critical ($X^* \approx 11$-13 nm). These findings are also supported by the results from the dynamics of MT lattice remodeling, which show that MT restructuring is 100% complete ($\chi \approx 1$) over 10-20 ms for $X$ = 7 and 11 nm indentation, but is incomplete ($\chi \approx 0.90$-0.93) for $X$ = 21 nm indentation (*the bottom inset* in Figs. 2, S1). This demonstrates that ruptured lateral contacts between the adjacent protofilaments can be efficiently restored over the timescale of a few tens of microseconds. The disruption of longitudinal bonds, however, inflicts irreparable damage on the MT lattice.

**SOP model predicts strong interactions at the longitudinal and lateral tubulin interfaces:** Our finding that the MT lattice structure in the collapsed state is ~85-90% similar to the native state (Figs. 3 and S2) strongly suggests that the compression-induced alterations in the MT lattice are mostly localized to the lateral and longitudinal interfaces, consistent with recent results from other groups.[41] This property allowed us to probe the thermodynamics of MT deformation. We analyzed the *FX* curves for forced indentation and force-quenched retraction (Supporting Movie S2) to determine the enthalpy change $\Delta H$, reversible work $w_{rev}$ and free energy change $\Delta G$ for the MT transitioning from the native state ($X = 0$) to the collapsed state ($X = 20$ nm) (see SI for more detail). Next, we calculated the enthalpy change and free energy change for disruption of the contacts at the lateral interface ($\Delta H_{lat}$ and $\Delta G_{lat}$) and longitudinal interface ($\Delta H_{long}$ and $\Delta G_{long}$). The number of lateral/longitudinal contacts was determined using a conservative estimate of the distance characteristic of contact disruption (see SI). A comparison of $\Delta H_{lat}$, $\Delta H_{long}$, and $\Delta G_{lat}$, $\Delta G_{long}$ demonstrates that these state functions show little variation with tip size and tip position on the MT (Table S2). The change in enthalpy, free energy, and entropy for the disruption of one lateral bond and one longitudinal bond are summarized in Table 2. Importantly, the obtained values of $\Delta G_{lat} = 6.9 \pm 0.4$ kcal/mol and $\Delta G_{long} = 14.9 \pm 1.5$ kcal/mol indicate that the intra-protofilament longitudinal tubulin-tubulin bonds are roughly twice as strong as the lateral inter-protofilament tubulin-tubulin bonds. This is consistent with our finding that the lateral bonds dissociate prior to the longitudinal bonds. Interestingly, the difference in entropy for the rupture of longitudinal bonds vs. lateral bonds is roughly four-fold (Table 2). This large $T\Delta S$ difference can be understood from the increased flexibility of the newly created protofilament ends. We also estimated the range for tubulin-tubulin interactions (see SI), and found that the longitudinal bonds are characterized by the longer interaction range ($\Delta y_{long} \approx 1.25\text{-}1.5$ nm) compared to the lateral bonds ($\Delta y_{lat} \approx 0.85\text{-}1.1$ nm). Both of these values lie within the 1.5-2 nm interaction range characteristic of protein complexes.[42,43]

**Nanoindentation spectra of a single protofilament suggest that it has large flexural rigidity:** We analyzed mechanical deformations of the MT cylinder using a thin-shell approximation (see SI).[32] For the average slope of the *FX* curves of $K_{MT} = 51.8 \pm 2.8$ pN/nm (for simulations with 10 nm tip; Table 1), the MT flexural rigidity *EI* comes to $(25,400 \pm 1,500) \times 10^3$ pN nm$^2$, which corresponds to the MT persistence length $L_p = EI/k_BT$ of $6.18 \pm 0.36$ mm. To estimate the flexural rigidity of tubulin strands, we performed simulations of bending deformation of single protofilaments formed by 8 (PF8/1), 16 (PF16/1), 24 (PF24/1), and 32 (PF32/1) dimers (see Supporting Movie S3 for PF16/1). In these simulations, the protofilament ends were clamped and the bending in response to forced indentations was examined. The simulated force-deformation spectra, i.e. the profiles of the deformation force $F$ vs. the cantilever tip displacement (deformation) $X$ (the $FX$ curves), and the corresponding profiles of the deformation energy (obtained by calculating the area under the $FX$ curve) vs. $X$ are presented in Fig. S5. Using the harmonic approximation valid for small 2-3 nm deformations we find that the values of $EI$ for these protofilament fragments are in the range of 18,000–26,000 pN nm$^2$ (Fig. S5, Table 3), which corresponds to the 4.5-6.6 µm range for the persistence length (Table 3).

## DISCUSSION

**A novel approach to multiscale modeling of MT polymer:** Dynamic, mechanical and force-generating properties of MTs play important roles in many cellular processes, but little is known about the thermodynamics of tubulin-tubulin interactions and mechanics of individual protofilaments that form the MT lattice. Previously, the energies of lateral and longitudinal bonds' dissociations were estimated with the help of molecular-mechanical models, in which the tubulin monomer/dimer was the smallest unit.[19-21] The major drawback of this approach is that tubulin energies are derived from the dynamic parameters of MT assembly and disassembly, which report on the thermodynamics of tubulin-tubulin interactions only indirectly. In contrast, the AFM-based dynamic force measurements provide a more straightforward experimental avenue, because in these experiments the protofilaments' deformation and tubulin-tubulin bonds rupture events are recorded with high spatial and temporal resolution.[32,33] However, due to the complexity of the multi-protofilament MT structure, the molecular interpretation of experimental force-indentation spectra at the level of protein-protein bonds is not trivial, as it requires the structure-based understanding of fine features of the experimental spectra.

We have overcome this limitation by carrying out the dynamic force measurements *in silico* using the amino acid as the smallest structural unit. Our computer-based experiments mimic the AFM based dynamic force experiments *in vitro*. The full control over the system we have during the entire course of forced deformation (contact point and direction of force application, constrained residues, indenter size) and the structural resolution (intact versus disrupted lateral/longitudinal interfaces) allows us to directly correlate the energy changes with the structure alterations at the residue level. Our approach to forced indentation *in silico* involves following stochastic dynamics of mechanical deformation of a biological particle, which is microscopically reversible when a force loading is sufficiently slow. In this regime of compressive force application, the rate of force increase is slower than the rate of system re-equilibration at each point along the deformation reaction path (quasi-equilibrium). This can be gleaned, e.g., from the comparison of *FX* curves for the 24-dimer long protofilament fragment PF24/1 obtained using varying cantilever velocities (see Fig. S6). We see that as $v_f$ decreases, the *FX* curves become less and less different. For example, the *FX* curves obtained for PF24/1 with $v_f$ = 1.0 and 0.5 μm/s look almost identical implying similar mechanical responses (Fig. S6). These results show that *in silico* indentation experiments reported here are carried out under near-equilibrium force-loading conditions.

The dynamic force spectroscopy *in silico* was previously applied by us to examine the forced unfolding of fibrin polymers[44] and to map the free energy landscape for deformation of the Cowpea Chlorotic Mottle Virus capsid.[45] This approach is made possible by combining the atomic-level and $C_\alpha$-based coarse-grained modeling with nanomanipulation of the MT lattice *in silico*.[34,35] By taking advantage of the computational acceleration on a GPU we were able to carry out detailed exploration of MT biomechanics on a long timescale (~50 ms) using the experimentally relevant conditions of force application.[32,33] With this approach, the atomic-level details underlying the lateral and longitudinal interactions are implicit in the SOP model of the MT cylinder. The next-neighbor interactions that stabilize MT structure and the lattice confinement for individual dimers are explicitly described. Within the context of these advances, although this model can be applied to describe processes that occur on a millisecond timescale, given current computational limitations it cannot be applied to much slower processes, such as the rupture of tubulin bonds and protofilament bending during MT depolymerization.

Here, we have determined the thermodynamic and mechanical characteristics of the MT that are difficult to access experimentally by analyzing the force-deformation curves from *in silico* nanoindentation simulations. The area under the *FX* curve is the total work performed on the system, and the reversible part of work can be linked to the Gibbs free energy change. Hence, the obtained free energies for the tubulin bonds' dissociation are based on theoretical analyses of *in silico* experiments, in which these bonds are directly manipulated. The only free parameter of the SOP model is $\varepsilon_h$, but the values of this quantity for each group of contacts between amino acids were calculated using the all-atom MD simulations (Table S3). The SOP model based nanoindentation assays provided the force-indentation curves which agree well with the experimental AFM spectra.[32,33] Thus, the agreement between the experiment and simulations was achieved without model fitting and without adjustment of free parameters. The detailed understanding of the mechanisms of MT deformation and structural collapse that we have achieved, offers unique insights into the mechano-chemistry of the MT lattice.

**Insights into the thermodynamics of tubulin-tubulin interactions in the MT lattice:** First, we found that the compressed MT behaves as a system of rigid elements interconnected through a network of lateral and longitudinal elastic bonds. Large rigidity of tubulin monomers agrees well with the results of prior computational modeling study, in which tubulin monomers were found to have a stable central core.[46] Accordingly, under small deformations the MT cylinder responds elastically, while undergoing continuous deformation characteristic of long wavelength modes. The importance of slow global modes in mechanical deformation of the MT lattice has also been revealed in a recent elastic network modeling study.[47] Beyond 3-4 nm compression, the αβ-tubulin dimers buckle, leading to flattening of the MT cylinder. With further compression the lateral contacts between the adjacent protofilaments dissociate (both α-α and β-β lateral bonds behave similarly), which is then followed by the rupture of longitudinal inter-dimer but not intra-dimer bonds (Fig. S4); these discrete structure changes are characteristic of short wavelength modes. Importantly, the sequence of microscopic events during mechanical MT compression defined here is likely to provide a blueprint for a pathway of normal MT disassembly. Indeed, in the course of mechanical compression tubulin bonds dissociate in the same order as during normal MT disassembly (lateral bonds prior to the longitudinal ones). Since the Gibbs free energy change for any bond dissociation is a state function, and therefore, does not depend on the exact cause of transition, our findings from nanoindentation experiments are directly applicable to describe and model the molecular events during MT disassembly.

Second, our work provides direct estimates of the free energies of dissociation of the lateral and longitudinal tubulin-tubulin bonds. We show that interfaces between tubulin subunits in the MT wall are characterized by strong non-covalent interactions. Structural analyses revealed that i) the α-α and β-β lateral interfaces are formed by a total of 19 and 21 stable residue-residue contacts, respectively, and that ii) the longitudinal intra-dimer bonds and inter-dimer bonds are stabilized by 78 and 38 total contacts, respectively. Both in the lateral and longitudinal interfaces, the most stable residue-residue contacts are hydrophobic bonds and salt bridges (Fig. S4A-D), consistent with a recent molecular modeling study[48] (major structural determinants in tubulin monomers involved in the inter-monomer contacts' formation are accumulated in Table S4). The difference between the obtained energies of the lateral and longitudinal bonds, $\Delta\Delta G = \Delta G_{long} - \Delta G_{lat} = 8.0$ kcal/mol (~13.3 $k_BT$), is close to the 11 $k_BT$ estimate reported earlier.[15,19] Importantly, the large strength of the tubulin-tubulin bonds reported here should prompt a re-evaluation of current molecular models of MT dynamics and stability, where these characteristics play a significant role.

Third, our results reveal that the rupture of the lateral tubulin-tubulin bond is highly reversible. This finding provides an important clue for understanding the molecular mechanisms of MT rescue, an abrupt switch from MT depolymerization to polymerization. We suggest that high reversibility of lateral bonds dissociation can promote "sealing" of the cracks between shortening protofilaments in the MT wall, thereby inhibiting the depolymerization. Another insight from our study concerns the geometry of contacts in the lateral tubulin interfaces. Previously, it has been suggested that the step-like "gaps" in the force-indentation curves obtained with AFM reflect the existence of two additional interaction sites between the dimers in adjacent protofilaments.[22] Our results demonstrate that a single pair of lateral interaction sites can account for all features of the experimental force spectrum, and the step-like gaps appear to result from the dissociation of lateral bonds under 6–8 nm deformations (Fig. 4). Interestingly, subjecting the MT lattice to mechanical stress can lead to the formation of small defects at the junction points connecting α- and β-tubulins; these defects grow in size with increasing force load (Fig. S3). Such defects could also emerge due to the mechanical "fatigue" of MT lattice that repeatedly experience large deformations, similar to the behavior of carbon nanotubes.[49,50] However, since the local force to cause such a crack is large (>300 pN), thermal vibrations of the MTs are highly unlikely to lead to the MT lattice fatigue,[33] so they cannot explain the MT "aging" *in vitro*.[51] Yet, crack formation might be pertinent to the activity of MT-severing enzymes, which are thought to exert large local forces while pulling the tubulins out of the MT wall.[52]

We found that the rupture of the longitudinal tubulin-tubulin bond is irreversible on the timescale of a few tens of microseconds. The corresponding irreversibility of the microtubule (MT) lattice restructuring is directly related to the long timescale required for the re-formation of all the longitudinal bonds and recovery of the MT lattice structure upon large indentation, a feature that was also detected in the original AFM experiments.[32,33] The authors found that it would take ~4 min for the MT lattice to fully self-heal (in the absence of free tubulin in solution) following an indentation beyond 10 nm. Hence, this finding shows that even after disruption of lateral and longitudinal bonds the lattice is able to recover, but only on a long timescale of a few minutes. To further explore this (ir)reversibility aspect, we carried out simulations of force-induced forward indentation followed by force-quenched backward tip retraction, but for the short protofilament fragment PF8/1. The results are presented in Fig. S7. Interestingly, we found that following the initial dissociation of the longitudinal bond between the 4-th and 5-th dimers, when the direction of tip motion was reversed the bond re-formed and the protofilament unbent completely over a 10-20 ms timescale (Fig. S7). Hence, the longitudinal bond dissociation in short protofilament fragments (such as PF8/1) is fully reversible on the millisecond timescale. This can be explained by lower entropic barriers for re-association of tubulin dimers in single protofilaments compared to the MT lattice.

**Implications for the models of force generation by the depolymerizing MT:** The large flexural rigidity of individual protofilaments reported here implies that tubulin protofilaments are fairly rigid. Our estimates (18,000–26,000 pN nm$^2$, Table 3) come close to the experimental values from several studies (Table 1 in Ref. (11)) and agree with those from a recent all-atom MD simulation study ($EI$ = 27,740 pN nm$^2$)[28]. This large protofilament's rigidity supports the idea that almost the entire energy from GTP hydrolysis is stored as mechanical tension in a straightened protofilament.[53,54] The obtained values of the persistence length for the MT cylinder $L_p$ = 6.18 mm agrees well with the experimentally measured value 3.45 mm (Ref. 1) and with the estimates from other computation studies.[28,47,55], demonstrating the validity of SOP modeling. Importantly, our results imply that tubulin protofilaments are about three orders of magnitude more flexible than the intact MT cylinder (of comparable length), which is in tune with some but not all previous estimates.[27,48,56] We also obtained a similar three order of magnitude difference for the persistence length, i.e. 6.18 mm for the MT cylinder vs. 4.5-6.6 μm for the tubulin protofilaments (Table 3), implying a stabilizing role played by the lateral tubulin-tubulin bonds.

Our work has important implications for the mechanism of force generation by the disassembling MTs. The ability of shortening MTs to transport a large cargo *in vitro* (up to 30 pN)[30] has been proposed to result from different mechanisms, including the biased-diffusion and power stroke-based models.[10,27] Only the latter mechanism takes a direct advantage of the tubulin dissociation pathway during which the lateral tubulin bonds dissociate prior to the longitudinal ones, causing a splaying of the protofilaments into the "ram's horns" structures.[14,58] This structural transition has been proposed to exert a power-stroke, capable of moving the appropriately attached cargo.[21,57,59] Although SOP modeling does not allow one to calculate directly the MT disassembly due to the prohibitively large computational time, it is interesting to discuss the estimates obtained here in the framework of the current models for MT force generation. The work to straighten the 8-32-dimer long protofilament with the intra-dimer bending angle of 0.4 rad can be estimated from the protofilament's rigidity, assuming that the protofilament behaves as a Hookean spring: ~39 –58 $k_BT$. This large bending energy suggests that a significant portion of this chemical energy can be converted into useful work[53,54]. Therefore, both the large flexural rigidity of tubulin protofilaments and high tubulin-tubulin dissociation energies obtained here support the proposal that the disassembling MT can serve as a strong depolymerization motor.[10,21] We hope that future advances in computational molecular modeling and availability of high-resolution, nucleotide specific tubulin structures will enable a direct testing of these conclusions. The modeling tools we have developed here can also be applied to study other complex biological assemblies when their detailed physico-chemical characteristics cannot be resolved using modern experimental approaches.

**MATERIALS AND METHODS**

**Computer model of MT lattice:** The structure of a finite-length fragment of the MT lattice was obtained from the 13 subunit ring structure of αβ-tubulin dimers, as in Ref. (60). This ring structure utilizes atomic coordinates of the αβ-tubulin dimer (PDB code: 1JFF),[61] in which the E-site in β-tubulin contains GDP. The finite-length fragment of GDP-tubulin MT lattice (MT8/13, Fig. 1A) was constructed by replicating the ring structure 8 times using the shift distance of 85 Å to obtain an MT construct of 8 dimers in length (MT8/13; see Fig. 1B). We used the all-atom Molecular Dynamics simulations in implicit water (SASA and GB models of implicit solvation) to obtain an accurate parameterization of the SOP model, as described below (see Table S3 in SI). The structures of 8, 16, 24, and 32 dimer long single protofilaments (PF8/1, PF16/1, PF24/1, and PF32/1) were extracted and replicated starting from the structure of MT8/13.

**Self-Organized Polymer (SOP) model:** We used the SOP model of the polypeptide chain[36] to describe each monomer (α-tubulin and β-tubulin). In the topology-based SOP model, each amino acid is represented by a single interaction center ($C_\alpha$-atom), and the $C_\alpha$-$C_\alpha$ covalent bond with the bond distance $a = 3.8$ Å (peptide bond length). The potential energy function of the protein conformation $U_{SOP}$ specified in terms of the coordinates $\{r_i\} = r_1, r_2, ..., r_N$ ($N$ is the total number of amino acid residues) is given by $U_{SOP} = U_{FENE} + U_{NB}^{ATT} + U_{NB}^{REP}$. The finite extensible nonlinear elastic potential $U_{FENE} = -\sum_{i=1}^{N-1} \frac{k}{2} R_0^2 \log\left(1 - \frac{(r_{i,i+1} - r_{i,i+1}^0)^2}{R_0^2}\right)$ with the spring constant $k = 14$ N/m and the tolerance in the change of a covalent bond distance $R_0 = 2$ Å describes the backbone chain connectivity. The distance between residues $i$ and $i+1$, is $r_{i,i+1}$, and $r^0_{i,i+1}$ is its value in the native (PDB) structure. We used the Lennard-Jones potential $U_{NB}^{ATT} = \sum_{i=1}^{N-3}\sum_{j=i+3}^{N} \varepsilon_h \left[(r_{ij}^0/r_{ij})^{12} - 2(r_{ij}^0/r_{ij})^6\right] \Delta_{ij}$ to account for the non-covalent (non-bonded attractive) interactions that stabilize the native folded state. We assumed that if the non-covalently linked residues $i$ and $j$ ($|i-j| > 2$) are within the cut-off distance $R_C = 8$ Å in the native state, then $\Delta_{ij} = 1$, and is zero otherwise. The value of $\varepsilon_h$ quantifies the strength of the non-bonded interactions. The non-native (non-bonded repulsive) interactions $U_{NB}^{REP} = \sum_{i=1}^{N-2} \varepsilon_l (\sigma_l/r_{i,i+1})^6 + \sum_{i=1}^{N-3}\sum_{j=i+3}^{N} \varepsilon_l (r_{ij}^0/r_{ij})^6 (1 - \Delta_{ij})$ are treated as repulsive. An additional constraint was imposed on the bond angle formed by residues $i$, $i+1$, and $i+2$ by including the repulsive potential with parameters $\varepsilon_l = 1$ kcal/mol and $\sigma_l = 3.8$ Å. These determine the strength and the range of the repulsion.

**Parameterization of $C_\alpha$-based SOP model:** A more detailed description of the SOP model parameterization is presented in the SI. In short, in the SOP model, the parameter $\varepsilon_h$ defines the average strength of non-covalent residue-residue contacts that stabilize the native state. Importantly, the values of $\varepsilon_h$ were calculated directly using MD simulations of an atomic structure model MT8/13 of the MT lattice at $T = 300$ K. The atomic-level details that determine the type and number of binary contacts between amino acids and their energies were ported to the SOP model of the MT lattice. Three 10 ns simulation runs were performed to calculate for each group of contacts the average non-bonded energy ($E_{nb}$), given by the sum of the van-der-Waals energy (Lennard-Jones potential) and the electrostatic energy (Coulomb potential), and the average number of binary contacts between amino acids ($N_{nb}$) that stabilize the native MT structure (native contacts). We assumed that a pair of residues formed a contact if the distance between their $C_\alpha$-atoms in the native state does not exceed the cut-off distance $R_C$. We used a standard choice of the cut-off distance $R_C = 8$ Å. All the native contacts were divided into five groups (contact types): (1) the intra-monomer contacts in the α-tubulin monomers; (2) the intra-monomer contacts in the β-tubulin monomers; (3) the intra-dimer contacts that stabilize dimer's structure; (4) the longitudinal inter-dimer contacts between any two dimers along the MT cylinder axis; and (5) the lateral inter-dimer contacts between the α-tubulin monomers and between the β-tubulin monomers in adjacent protofilaments. To calculate the energy for non-bonded interactions, we employed the Solvent Accessible Surface Area (SASA)[62] and Generalized Born (GB)[63] models of implicit solvation, which are based on the CHARMM19 force-field.[64] We used the output from SASA model based simulations (coordinate and energy files) to calculate the values of $E_{nb}$ and $N_{nb}$ for the contact groups 1-3. Since electrostatic interactions are important for the formation of longitudinal and lateral tubulin-tubulin bonds, we used a more accurate GB model to calculate $E_{nb}$ and $N_{nb}$ for the contact groups 4 and 5. Finally, dividing $E_{nb}$ by the corresponding value of $N_{nb}$ for each contact group we obtained the value of $\varepsilon_h$ (see Table S3), which were used in all simulations reported here.

**Dynamic force measurements *in silico*:** Nanoindentation measurements were performed at different points on the MT surface using the spherical tip of radius $R_{tip} = 10$ and 15 nm (Fig. 1C), similar to the 15-nm tip used in atomic force microscopy (AFM) experiments.[32,33] In the simulations of mechanical indentation of the MT cylinder and deformation of single protofilament fragments, the tip-MT lattice interactions and the tip-protofilament interactions were modeled by the repulsive potential, $U_{tip} = \varepsilon_{tip}[\sigma_{tip}/(r_i - R_{tip})]^6$, where $r_i$ is the position of the *i*-th particle, $\varepsilon_{tip} = 4.18$ kJ/mol, and $\sigma_{tip} = 1.0$ Å. In the forward indentation measurement, the tip exerted the time-dependent compressive force $f(t) = f(t)\mathbf{n}$ in the direction $\mathbf{n}$ perpendicular to the MT or protofilament surface (Fig. 1B). The force magnitude $f(t) = r_f t$ was increased with the force-loading rate $r_f = \kappa v_f$, where $v_f = 1.0$ µm/s (for MT indentations) and $v_f = 0.2$ µm/s (for protofilament deformations) is the velocity of the cantilever base (piezo) represented by a virtual particle, and $\kappa = 0.05$ N/m is the cantilever spring constant. The force $f(t)$ is transmitted to the tip through the cantilever spring, and the resisting indentation force (for MT lattice) or deformation force (for the protofilament) $F$ is calculated using the energy output from simulations. In the simulations of backward (tip) retraction, the direction of tip (or piezo) motion was reversed, which resulted in a gradual decrease of the compressive force to zero.

*Simulations of mechanical indentation of MT in silico* were performed using the SOP model of MT8/13 and Langevin simulations accelerated on a GPU[34,35]. To account for the long persistence length of MTs (microns to millimeters),[65] positions of the $C_\alpha$-atoms for a total of 9 residues 248, 253, 257, 262, 325, 326, 329, 348, 349 in each tubulin monomer at the plus MT end, and positions of the $C_\alpha$-atoms for 9 residues 98, 176, 177, 180, 221, 224, 225, 403, 407 in each tubulin monomer at the plus MT end were constrained (Fig. 1). We implemented hard constraint conditions, in which the positions of all 9 constrained residues were fixed, and soft harmonic constraints, in which we connected these same 9 residues to a virtual wall through a harmonic spring with the spring constant of $\kappa = 0.1$ nN/nm. A total of 42 indentation runs were generated (using hard constraints): 3 runs per indentation point 1-7; 21 runs for each $R_{tip}$ value. We profiled the dependence of the indentation force $F$ in dynamic force experiments on the cantilever tip displacement $X$ (indentation depth) and $Z$ (the piezo displacement) for all indentation points (Fig. 1C). The *FX* curves show higher sensitivity to the MT deformation dynamics than *FZ* curves, and so $X$ is a better reaction coordinate. However, for the purpose of comparing the results of experiments with simulations, we analyzed the *FZ* and *FX* curves (Fig. 2A).

*Simulations of bending deformation of MT protofilaments in silico* were performed using the SOP models of protofilament fragments PF8/1, PF16/1, PF24/1, and PF32/1, and Langevin simulations on a GPU. To constrain a protofilament, we fixed same positions of the $C_\alpha$-atoms at the N-terminus of the first monomer and at the C-terminus of the last monomer. The cantilever tip was set to move in the direction perpendicular to the protofilament axis, as shown in Fig. S5. Total of 12 indentation runs for indentation point 3 were generated for all four protofilament fragments: 3 runs per fragment ($R_{tip} = 10$ nm). We profiled the dependence of the deformation force $F$ on the cantilever tip displacement $X$ for the protofilament fragments (Fig. S5). Flexural rigidity of the protofilament fragments was calculated in the harmonic approximation as described in SI. Because the harmonic approximation valid only in the regime of small deformations, we analyzed the initial quadratic 2-3 nm portion of the *FX* curves (Fig. S5). This deformation translates to the average dimer-dimer bending angle of $1-2°$.

*Free energy estimation:* To obtain accurate estimates of the Gibbs free energy change associated with the disruption of single lateral interface and longitudinal interface ($\Delta G_{lat}$ and $\Delta G_{long}$), we determined the mechanical work performed on the MT (area under the *FX* curve). Since it is not possible to apply infinitely slow force loading, which would correspond to the equilibrium conditions of mechanical force application (and reversible work), the total work ($w$) in our indentation cycle contains the reversible part $w_{rev}$, which is spent to deform the MT lattice, and to dissociate the lateral and longitudinal bonds, as well as the irreversible part $w_{irrev}$ (see hysteresis in the *FX* curves in Figures 2 and 3; see also Figures S1 and S2 in SI). We calculated the reversible part of work $w_{rev}$ using the Crooks theorem (see SI).

**Analyses of the simulation output** for mechanical compression of MT8/13, and bending deformation of PF8/1, PF16/1, PF24/1, and PF32/1, including the structure visualization and determination of the thermodynamic quantities ($\Delta G$, $\Delta H$, and $T\Delta S$), flexural rigidity ($EI$), persistence length ($L_p$), and the range of lateral and longitudinal bonds ($\Delta y$) are described in detail in SI.

**Associated content:** *Supporting Materials and Methods*: additional description of multiscale MT modeling approach, construction of MT model, main features of SOP model, its parameterization and dynamic force measurements *in silico*, and methods for data analysis; *Supporting Tables:* summary of microscopic transitions in the MT lattice, comparison of the mechanical and thermodynamic parameters from indentations at different points, and parameterization of the SOP model for the MT lattice; *Supporting Movies*: forced indentation of MT lattice, forward indentation and backward tip retraction of the MT lattice and force deformation of 16-dimers long MT protofilament; *Supporting Figures:* mechanism of MT lattice nanoindentation with $R_{tip}$ = 15 nm, detailed information about structure changes during force indentation, atomic representation of lateral and longitudinal tubulin-tubulin interfaces, force-deformation spectra for single protofilament fragments, dependence of force-deformation spectra on cantilever velocity, and reversibility of longitudinal bonds' dissociation observed for single protofilament fragment. This material is available free of charge via the Internet at http://pubs.acs.org.

**Acknowledgements:** We thank J.R. McIntosh for critical reading of the manuscript, A. Zaytsev for help with videos. This work was supported in part by National Institutes of Health Grant R01-GM098389 to ELG; by the National Science Foundation CAREER award MCB-0845002 to RID; by Russian Fund for Basic Research (13-04-40188-H, 13-04-40190-H) and Presidium of Russian Academy of Sciences (Mechanisms of the Molecular Systems Integration and Molecular and Cell Biology programs) to FIA.

**Table 1: Comparison of the mechanical properties of the MT lattice determined from indentations *in vitro* and *in silico*.** Values are means with standard deviations: MT spring constant – $K_{MT}$, critical force – $F^*$, and critical distance – $Z^*$ (cantilever velocity $v_f = 1.0$ µm/s). Experimental data *in vitro* are from the results of de Pablo et. al[32] and Shaap et. al[33] who used $v_f \approx 0.2$ µm/s and $R_{tip} \approx 15–20$ nm. The experimental values of $K_{MT}$ were extracted from the experimental histogram (Fig. 2 in Ref. (31)); the values of $F^*$ and $Z^*$ were taken from the experimental force-indentation curves (Fig. 1 in Ref. (33)). Theoretical values of $K_{MT}$, $F^*$, and $Z^*$ were obtained by averaging over 3 simulation runs for each indentation point 1-7.

| Indentation | $K_{MT}$, pN/nm | $F^*$, nN | $Z^*$, nm |
|---|---|---|---|
| *in silico* ($R_{tip}$ = 10 nm) | 51.8±2.8 | 0.62±0.07 | 23.8±2.2 |
| *in silico* ($R_{tip}$ = 15 nm) | 61.4±6.6 | 0.76±0.04 | 26.9±1.3 |
| *in vitro* ($R_{tip}$ = 20 nm) | 74.0±13.0 | 0.4±0.1 | 17.2±3.5 |

**Table 2: Thermodynamic characteristics deduced from *in silico* indentations**: change in enthalpy $\Delta H$, entropy $T\Delta S$ and free energy $\Delta G$ (with standard deviations) associated with the disruption of a single lateral bond (interface) and longitudinal bond. Averaging is performed over all indentation points (1-7, Fig. 1C) and for 10 and 15 nm tip.

| Interface | $\Delta G$, kcal/mol | $\Delta H$, kcal/mol | $T\Delta S$, kcal/mol |
|---|---|---|---|
| Lateral | 6.9±0.4 | 9.3±0.8 | 2.5±0.7 |
| Longitudinal | 14.9±1.5 | 25.7±2.2 | 10.8±2.5 |

**Table 3: Mechanical bending parameters deduced from *in silico* deformations.** Values are averages with standard deviations of the flexural rigidity ($EI$) and persistence length ($L_p$) obtained from 5 bending simulation runs for single protofilaments of 8 (PF8/1), 16 (PF16/1), 24 (PF24/1), and 32 (PF32/1) tubulin dimers.

| System | PF8/1 | PF16/1 | PF24/1 | PF32/1 |
|---|---|---|---|---|
| $EI \times 10^{-26}$, N m$^2$ | 1.8±0.1 | 2.3±0.2 | 2.6±0.1 | 2.6±0.1 |
| $L_p$, µm | 4.5±0.3 | 5.7±0.4 | 6.3±0.2 | 6.3±0.3 |

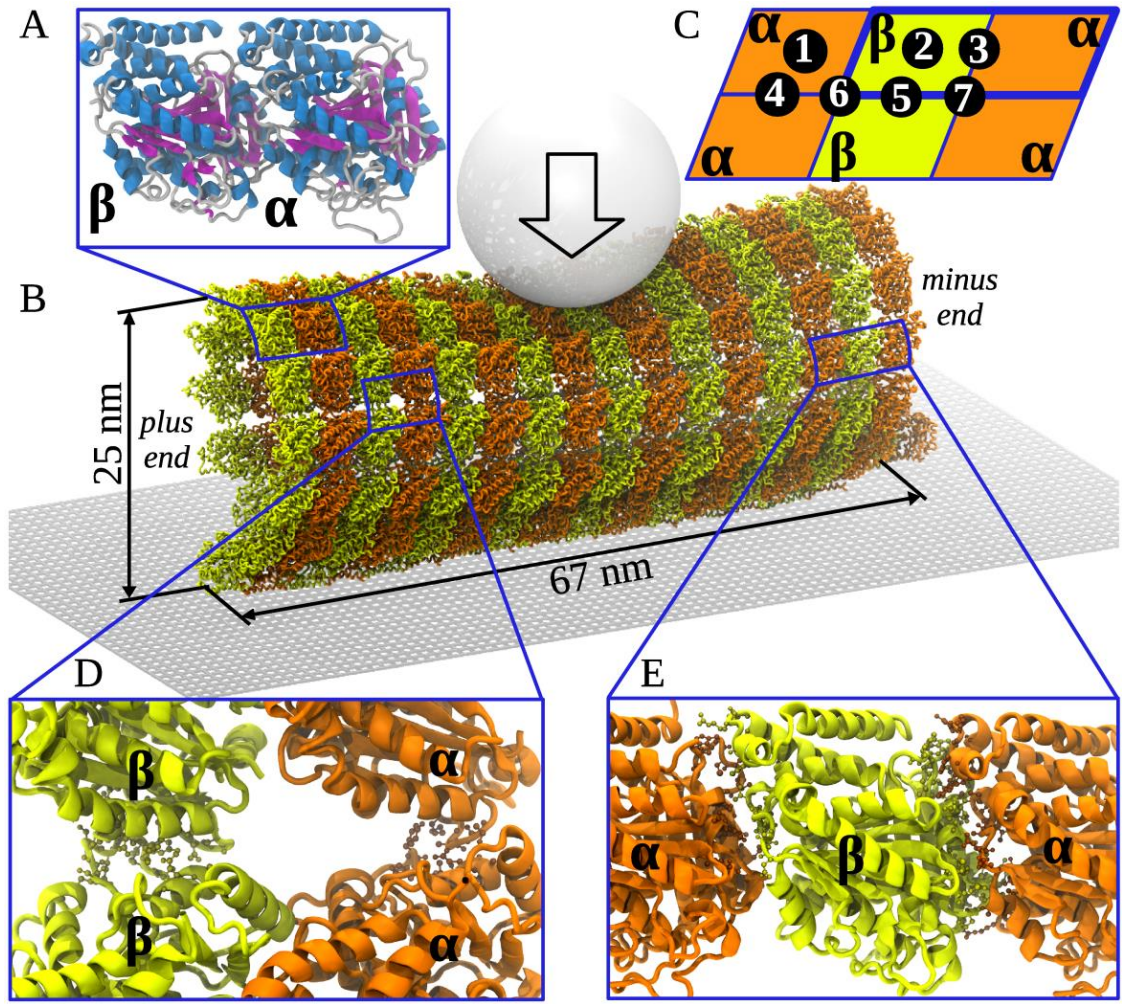

**Fig. 1. Schematic of SOP model for nanoindentations and subunit interactions. A:** Atomic model of αβ-tubulin heterodimer showing the secondary structure: helices (blue), sheets (red), and loops/turns (gray). **B:** The $C_\alpha$-based representation of an MT lattice fragment (MT8/13) with tubulin heterodimers shown in brown (α-tubulin) and green (β-tubulin). In simulations, the cantilever tip (gray ball) produces indentations of MT lattice (arrow shows direction of force). **C:** Schematic of MT fragment with locations of 7 specific points for indentation: for points 1, 2, and 3 the force is applied onto the surface of the α-tubulin, β-tubulin, and dimer-dimer interface, respectively; for points 4 and 5 force is applied at the interfaces between two α- and two β-tubulins, respectively; for points 6 and 7 force is applied between protofilaments at the junction connecting four tubulin dimers, and within two dimers, respectively. **D.** Atomic structure of the lateral (α-α and β-β) interfaces between adjacent protofilaments. **E.** Atomic structure of the longitudinal interfaces: inter-dimer (α-β; left) and intra-dimer (β-α, right).

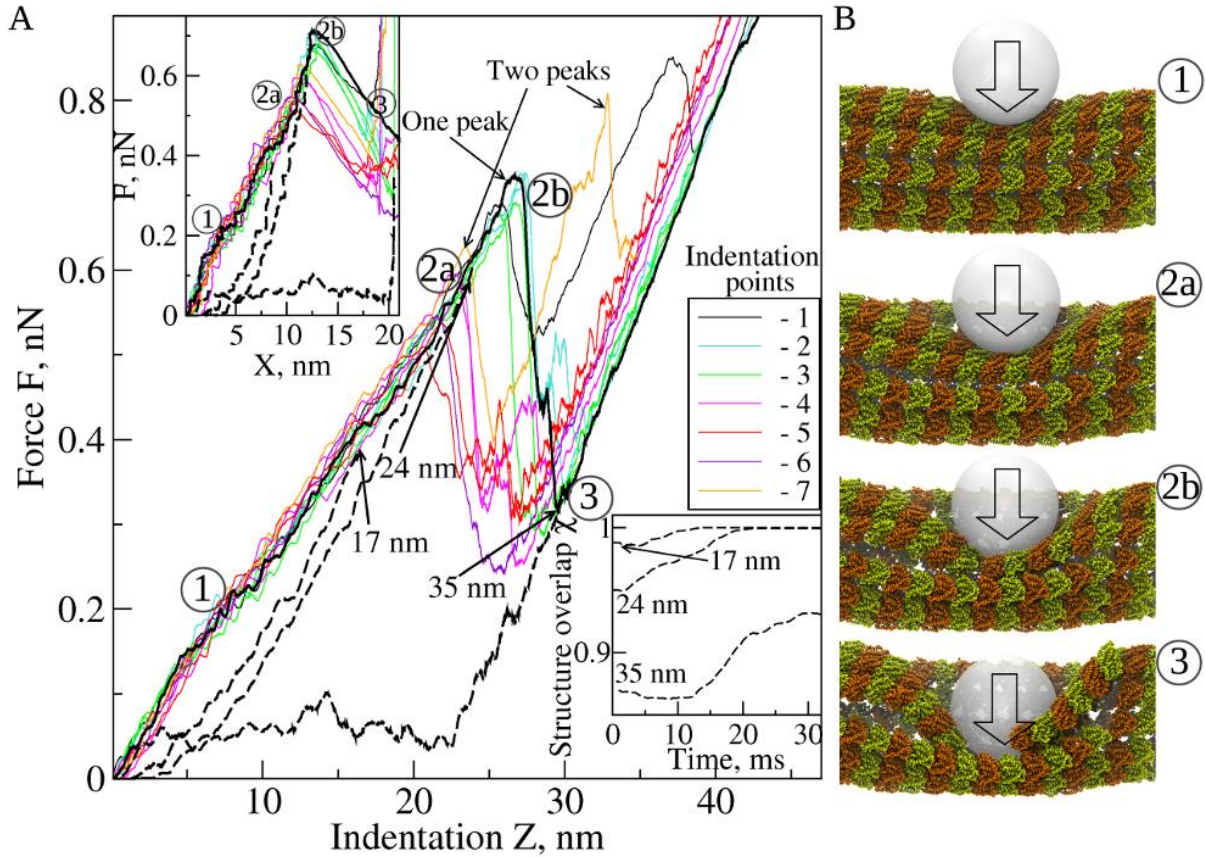

**Fig. 2. Force indentations *in silico*. A:** The force-deformation (*FZ*) curves for 7 indentation points (Fig. 1C), each depicted with different color. Results were obtained with $v_f = 1.0$ μm/s and $R_{tip}$ = 10 nm (see Fig. S1 for $R_{tip}$ = 15 nm). *Z* is the displacement of the virtual cantilever base (piezo in AFM). Dashed black curves represent the *FZ* profiles for the tip retraction simulations, which followed the forward indentations (solid black curve) with *Z* = 17, 24 and 35 nm as initial conditions. *The top inset* shows the corresponding *FX* curves for the forward indentation (solid curves; colors as in A) and backward tip retraction (dashed black curves). *X* is the displacement of the cantilever tip. *The bottom inset* shows the time profiles of the structure overlap χ for MT lattice restructuring during tip retraction (starting from *Z* = 17, 24 and 35 nm indentations). **B:** The MT structure snapshots 1, 2a, 2b, and 3 illustrating the mechanism of MT deformation and collapse. Structure 1: continuous deformation (*Z* < 15-20 nm; elastic regime). Structures 2a and 2b: disruption of lateral and longitudinal interfaces, respectively (20-25 nm < *Z* < 25-30 nm; transition regime). Structure 3: post-collapse evolution (*Z* > 25-30 nm). These structures correspond to the accordingly numbered regions in *FZ* and *FX* curves in **A**.

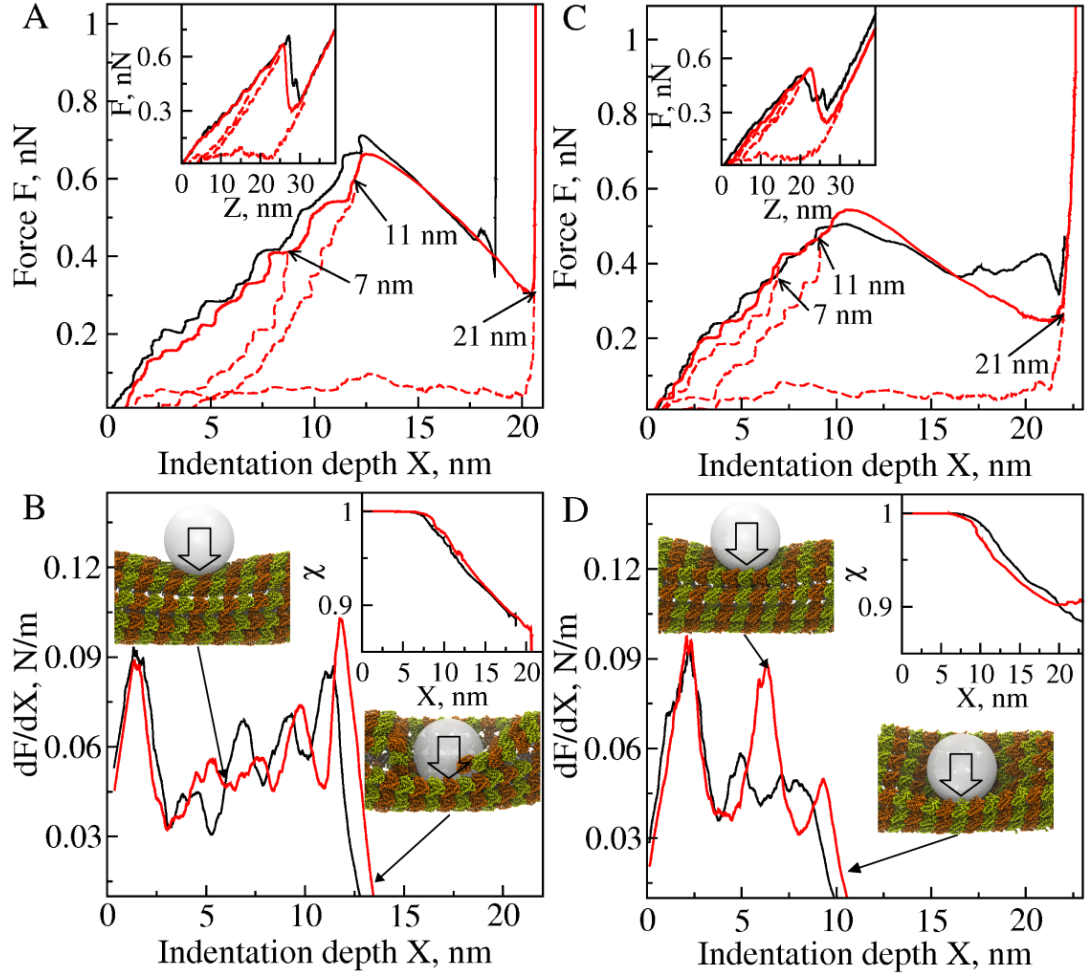

**Fig. 3. Force spectra for indentation and retraction and deformation-induced MT structure alterations.** Shown are results for indentation points 2 (black) and 3 (red) in **A** and **B**, and for indentation points 7 (black) and 6 (red) in **C** and **D** obtained with $v_f = 1.0$ μm/s and $R_{tip} = 10$ nm (see Fig. S2 for results obtained with $R_{tip} = 15$ nm). **A** and **C:** The *FX* curves for forward indentation (solid black and red lines). *The inset*s show the *FZ* curves. Curves for backward tip retraction (dashed red lines) were generated using the structures obtained from forward indentation for $X = 7$, 11, and 21 nm (indicated on the graphs). **B** and **D:** The slope $dF/dX$ for force spectra from panels **A** and **C**. Snapshots show the side-views of the MT structure before dissociation of the lateral bonds and after dissociation of the longitudinal bonds. *The insets* show the profiles of structure overlap $\chi$ vs. $X$.

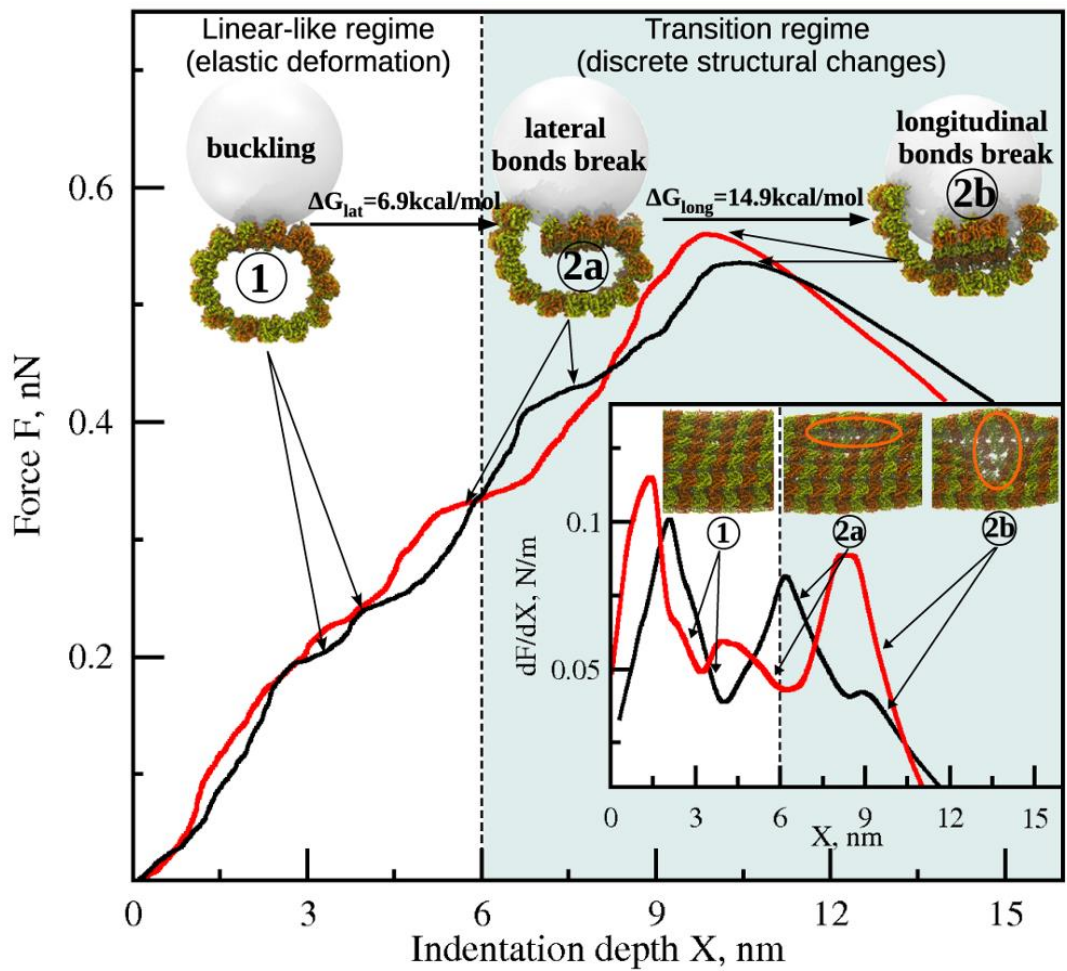

**Fig. 4. Summary of the MT deformation mechanics.** The *FX* curves for force application between protofilaments at the junction connecting four tubulin dimers (indentation point 6, red curve) and within the adjacent dimers (point 7, black curve) illustrate that the sequence of transitions is overall similar. Images of the MT lattice and tip on the main graph are shown in a MT cross-section view. White area indicates the linear-like regime for elastic deformations, represented by the buckled MT cylinder (structure 1, $X \approx 3\text{-}4$ nm). Gray area corresponds to the transition regime, characterized by the rupture of lateral (structure 2a, $X \approx 6\text{-}8$ nm), then longitudinal contacts (structure 2b, $X \approx 9\text{-}11$ nm). *The inset* shows the slope *dF/dX* vs. *X*, so the peaks correspond to the activated system states for the buckling and dissociation transitions. Images in the inset are the same structures 1, 2a, and 2b (with no tip) viewed from the top; areas circled in orange contain the disrupted lateral and longitudinal interfaces (see also Fig. S3).

**Supporting Information**

**Tubulin bond energies and microtubule biomechanics determined from nanoindentation *in silico***

Olga Kononova,[a,b] Yaroslav Kholodov,[b] Kelly E. Theisen,[c] Kenneth A. Marx,[a] Ruxandra I. Dima,[c] Fazly I. Ataullakhanov,[d,e,] Ekaterina L. Grishchuk,[f,*], and Valeri Barsegov[a,b,*]

[a]Department of Chemistry, University of Massachusetts, Lowell, MA 01854, USA; [b]Moscow Institute of Physics and Technology, Moscow region, 141700, Russia; [c]Department of Chemistry, University of Cincinnati, Cincinnati, OH 45221, USA; [d]Center for Theoretical Problems of Physicochemical Pharmacology, Russian Academy of Sciences, Moscow 119991, Russia; [e]Physics Department, Moscow State University, Moscow 119991, Russia; [f]Physiology Department, Perelman School of Medicine, University of Pennsylvania, Philadelphia, PA 19104

**Supporting Materials and Methods**

**Constructing the ring structure of αβ-tubulin dimers:** As a starting point for our computer model of a fragment of a MT cylinder we used the model from the published work of Wells and Aksimentiev.[1] To build the model, Wells and Aksimentiev used the all-atom Molecular Dynamics Flexible Fitting (MDFF) method to generate an atomistic structure for an MT lattice starting from the cryo-EM map of the lattice[2] and the atomistic structure of the tubulin subunit from X-ray crystallography (PDB entry: 1JFF).[3] The MDFF method locates tubulin subunits in the cryo-EM map by testing all plausible subunit locations by re-adjusting position of the atomic tubulin structure to achieve maximum correspondence to the cryo-EM data. Only standard potential energy functions (atomic force fields) implemented in all-atom Molecular Dynamics (MD) simulations were used without any simplifying assumptions about the shapes of these bond energies and without using the attractive potentials. Using this approach, two fragments were created: two protofilaments connected by lateral contacts as in the bulk of 13_3 MT lattice (system N), and two protofilaments connected by lateral contacts as seen at the seam in 13_3 structure (system S). By lateral contacts we mean a set of binary contacts between the amino acid residues forming the interfaces between adjacent tubulin monomers in neighboring protofilaments, as identified by the above MDFF procedure. Specifically, coordinates for one protofilament were obtained by Wells and Aksimentiev using rigid-fitting of the tubulin dimer into a subset of the cryo-EM map (CoLoRes program), and the best fit was saved. Coordinates for a second protofilament were obtained from the first protofilament by a rotation of $360°/13$ about the map's center, and an axial protofilament displacement of 3/2 dimer subunit. Then, the full cylindrical structure was built by assembling 12 copies of the N-type PFs and one copy of the S-type protofilament into $360°/13$ radial sectors. Next, the all-atom Molecular Dynamics simulations in explicit solvent (i.e. fully water-solvated protein system plus ions to neutralize charges on polypeptide chains) were used by Wells and Aksimentiev to determine the best possible configuration of all subunits in this system. This configuration corresponded to the minimum of the total (kinetic plus potential) energy of the system. No other kinetic or thermodynamic information has been used during the energy minimization step. All system elements were allowed to change their conformation to make the system capable of reaching the state of equilibrium with energy minimum. A similar approach has been used extensively to determine atomistic structures for large biomolecular complexes, including HIV-1 capsid structure,[4] the ribosome-SecYE complex,[5] the monomeric yeast and mammalian Sec61 complexes interacting with the translating ribosome,[6] the Trypanosoma brucei ribosome,[7] the yeast 26S proteasome,[8] and the no-go mRNA decay complex Dom34-Hbs1 bound to a stalled 80S ribosome.[9]

**Data analyses:**

1) *Structure visualization* was performed using the VMD package.[10] To quantify the extent of structural similarity between a given conformation and a native (reference) state, we used the structure overlap function, $\chi(t) = (2N(N-1))^{-1} \sum \Theta(|r_{ij}(t) - r_{ij}^0| - \beta r_{ij}^0)$, where in the Heaviside step function $r_{ij}(t)$ and $r_{ij}^0$ are the inter-particle distances between the $i$-th and $j$-th residues in the transient structure and in the native state, respectively, and $\beta = 0.2$ is the tolerance for the inter-particle distance change.

2) *Mechanical deformation of MT cylinder*: We used the thin shell approximation[11,12] to analyze the flexural rigidity of the MT cylinder, according to which the slope of the *FX* curve for small *X* is given by $K_{MT} = 1.18 E t^{5/2}/R^{3/2}$, where $t = 1.1$ nm is the cylinder thickness, $R = 12.5$ nm is the exterior radius, and $E$ is the Young's modulus. We assume that the MT cylinder is roughly isotropic prior to disruption of lateral contacts. The bending rigidity of the MT cylinder was estimated as the product $EI = E \times I$, where $I$ is the moment of inertia of the MT cylinder with respect to the longitudinal axis. Since the MT cylinder is composed of $n = 13$ identical protofilaments of radius $r = 2.5$ nm, the moment of inertia can be calculated as $I = (2/\pi^2 n^3 + n)\pi/4 r^4$. The persistence length was estimated by using the formula $L_p = EI/k_B T$.

3) *Bending deformation of protofilament fragments:* By performing numerical integration, the force-deformation (*F* vs. *X*) profiles obtained from bending simulations were transformed into the profiles of bending energy $V = \int F(X) dX$ vs. *X* (*the inset* in Fig. S5). To estimate the flexural rigidity for each protofilament, we calculated the energy of bending of the beam, $V = \frac{EI}{2} \int_L (1/R - 1/R_0)^2 dl$, where $R_0$ and $R$ are the radii of the curvature of the protofilament in the initial state, used as a reference structure, and the final state, respectively; $dl$ is the length element; and $L$ is the total length.[11] This is a harmonic approximation valid only for small deformations $X$. To find the range of $X$ for which this approximation is valid, we performed a fit of the quadratic function $V \sim X^2$ to the curves of the bending energy (dashed black line in *the inset* in Fig. S5). In the initial state, the protofilament is straight, and, hence, $1/R_0 = 0$. To simplify calculations, we only considered the protofilament portion of constant curvature, in which case $V = EIL/(2R^2)$. Knowing $V$ and $R$ allows us to calculate $EI = 2VR^2/L$. Using the simulation output, we estimated $R$ using the formula for the radius of the arc $R = X/2+C^2/(8X)$, where $C$ is the end-to-end distance (Fig. S5). The persistence length for each protofilament fragments was calculated using the formula: $L_p = EI/k_BT$.

4) *Estimation of $\Delta G_{lat}$ and $\Delta G_{long}$:* We analyzed the energy output from simulations (potential energy $U_{SOP}$) to estimate the enthalpy change of deformation $\Delta H$. The total work of deformation $w$ can be obtained by integrating the area under the *FX* curve for the forward indentation. This procedure can then be repeated for the retraction curve to resolve the reversible part of work $w_{rev}=\Delta G=\Delta H-T\Delta S$ and the entropic contribution $T\Delta S= \Delta H - w_{rev}$. We used this approach to resolve the free energy change associated with the rupture of lateral and longitudinal bonds. First, we carefully selected two short time intervals within the same trajectory, one showing the dissociation of lateral interfaces and the other showing the dissociation of longitudinal interfaces. Next, for each selected portion of the trajectory we calculated the number of dissociated lateral interfaces $n_{lat}$ and longitudinal interfaces $n_{long}$, and estimated the total reversible work $w_{lat}$ and $w_{long}$. The quantities $w_{lat}$ and $w_{long}$ were calculated by integrating the corresponding regions in the *FX* curve for the forward indentation and the backward retraction. The amount of dissipated energy (irreversible part of work $w_{irrev}$) was estimated using the Crooks theorem (see next paragraph). The free energy of dissociation per single lateral (longitudinal) bond was calculated by taking the ratio $\Delta G_{lat} = w_{lat}/ n_{lat}$ ($\Delta G_{long} = w_{long}/ n_{long}$).

5) *Calculation of reversible work $w_{rev}$:* Crooks showed theoretically[13] and the Bustamante lab verified experimentally (using single-molecule measurements on an RNA molecule[14]) that in a driven unfolding process from the initial point $X = X_{in}$ to some final point $X = X_{fin}$ and a reverse refolding process from $X = X_{fin}$ back to $X =X_{in}$, the probability distributions (histograms) of the values of work for the forward process $p_{unf}(w)$ and reverse process $p_{ref}(-w)$ are connected via the relationship:

$$p_{unf}(w) / p_{ref}(-w) = \exp\left[-(w - \Delta G_{eq})/k_B T\right]$$

The point at which $p_{unf}(w)$ and $p_{ref}(-w)$ intersect corresponds to the equilibrium work $w = \Delta G_{eq}$. In our simulations, we are limited to 3 trajectories for each indentation point 1-7 (Fig. 1C). We can still use the Crooks relationship if we notice that $p_{unf}(w)$ and $p_{ref}(-w)$ intersect at their tails, which correspond to the extreme observations: the minimum work ($w_{min,forw}$) for the forward process and the maximum work ($w_{max,rev}$) for the reverse process ($w_{max,ref} < w_{min,forw}$). Hence, the reversible work is between the minimum work for the forward process and the maximum work for the reverse process, i.e. $w_{max,rev} < w < w_{min,forw}$. In the context of our indentation experiments *in silico*, this corresponds to the minimum work $w_{min,ind} = min\{w_{1,ind}, w_{2,ind}, ...\}$ for the forward forced indentation process ($p_{ind}(w)$) and the maximum work $w_{max,ret} = max\{w_{1,ret}, w_{2,ret}, ...\}$ for the backward force-quenched retraction process ($p_{ret}(w)$), i.e. $w_{max,ret} < w < w_{min,ind}$. We estimated reversible work $w_{rev} = \Delta G_{eq}$ by taking the arithmetic mean of $w_{max,ret}$ and $w_{min,ind}$, i.e. $w \approx (w_{min,ind} + w_{max,ret})/2$.

6) *Estimation of the range of lateral and longitudinal bonds:* We examined the forced-induced flattening of the ring of tubulin dimers under the tip in the lateral direction $\Delta Y_{lat}$ and elongation of MT protofilaments in the longitudinal direction $\Delta Y_{long}$ (Fig. S3). We related the indentation depth $X$ to the resulting changes in MT-tip contact area along the cylinder axis and in the transverse direction. Corresponding to $X \approx 6\text{-}8$ nm indentation, when the lateral contacts become disrupted (force plateau in *FX* curves and structure 2a in Fig. 4), the elongation of the tubulin ring is $\Delta Y_{lat} \approx 3.3$ nm. From structure analysis we found the number of lateral interfaces under the tip to be $N_{lat} = 3\text{-}4$ (Fig. S3B, S3C). These were used to estimate the average extension per lateral interface at which the dissociation occurs, $\Delta y_{lat} = \Delta Y_{lat}/N_{lat} \approx 0.85\text{-}1.1$ nm. Next, the longitudinal contacts become disrupted at $X \approx 9\text{-}11$ nm indentation (force peak in the *FX* curves and structure 2b in Fig. 4), for which the elongation of 3-4 dimer-long portion of an MT protofilament under the tip is $\Delta Y_{long} \approx 7.5$ nm. Analysis of structures showed that the number of longitudinal interfaces involved was $N_{long} \approx 5\text{-}6$ (Fig. S3B, S3C) Hence, the average extension for dissociation of the longitudinal interface is $\Delta y_{long} = \Delta Y_{long}/N_{long} \approx 1.25\text{-}1.5$ nm.

**Supporting References:**

**Supporting Tables**

**Table S1: Summary of microscopic transitions**. These transitions accompany the mechanical compression and collapse of the MT lattice probed at seven unique indentation points 1-7 (Fig. 1C in main text).

| Indentation point | Summary of transitions |
|---|---|
| 1 | bending; disruption of 20 lateral interfaces; disruption of 5 longitudinal interfaces |
| 2 | bending; disruption of 24 lateral interfaces; disruption of 3 longitudinal interfaces |
| 3 | bending; disruption of 23 lateral interfaces; disruption of 4 longitudinal interfaces |
| 4 | bending; disruption of 16 lateral interfaces; disruption of 3 longitudinal interfaces |
| 5 | bending; disruption of 18 lateral interfaces; disruption of 3 longitudinal interfaces |
| 6 | bending; disruption of 19 lateral interfaces; disruption of 4 longitudinal interfaces |
| 7 | bending; disruption of 21 lateral interfaces; disruption of 4 longitudinal interfaces |

**Table S2: Comparison of the mechanical and thermodynamic parameters from indentations at different points on the MT lattice**. Simulations were carried out using $v_f = 1.0$ μm/s, and $R_{tip} = 10$ and 15 nm. Presented are the values of critical force $F^*$, critical indentation depth $X^*$, and spring constant $K_{MT}$. Also shown are the enthalpy change $\Delta H$, and free energy change $\Delta G$ associated with the dissociation of lateral and longitudinal bonds.

| Indentation point | $F^*$,nN | | $X^*$,nm | | $K_{MT}$,pN/nm | | $\Delta G_{lat}$, kcal/mol | | $\Delta G_{long}$, kcal/mol | | $\Delta H_{lat}$, kcal/mol | | $\Delta H_{long}$, kcal/mol | |
|---|---|---|---|---|---|---|---|---|---|---|---|---|---|---|
| | 10nm | 15nm | 10 nm | 15 nm | 10 nm | 15 nm | 10 nm | 15nm | 10 nm | 15 nm | 10 nm | 15 nm | 10 nm | 15 nm |
| 1 | 0.67 | 0.72 | 12.5 | 12.0 | 51 | 53 | 6.5 | 6.6 | 15.0 | 16.1 | 8.4 | 10.2 | 27.1 | 29.1 |
| 2 | 0.71 | 0.78 | 12.3 | 12.4 | 48 | 54 | 6.7 | 7.2 | 13.4 | 17.0 | 9.3 | 10.1 | 21.8 | 23.9 |
| 3 | 0.67 | 0.77 | 11.9 | 12.6 | 48 | 58 | 6.4 | 6.8 | 13.9 | 16.9 | 9.5 | 9.7 | 24.3 | 27.8 |
| 4 | 0.60 | 0.73 | 11.3 | 10.9 | 55 | 61 | 7.1 | 7.6 | 12.5 | 13.6 | 9.4 | 9.8 | 25.0 | 27.7 |

| 5 | 0.54 | 0.80 | 10.5 | 11.0 | 53 | 57 | 6.6 | 7.2 | 12.5 | 16.7 | 7.9 | 8.6 | 26.1 | 26.3 |
| 6 | 0.54 | 0.71 | 10.3 | 10.8 | 54 | 66 | 6.3 | 7.1 | 17.8 | 16.2 | 8.2 | 8.8 | 21.7 | 24.0 |
| 7 | 0.63 | 0.74 | 10.8 | 10.9 | 55 | 65 | 7.1 | 7.3 | 14.3 | 16.2 | 10.4 | 10.5 | 26.7 | 27.8 |

**Table S3: Parameterization of the SOP model for the MT lattice**. Summarized for each group of residue-residue contacts 1-5 are the average number of native contacts ($N_{nb}$), the average energy of non-covalent (non-bonded) interactions ($E_{nb}$), and the average strength of non-covalent interactions per contact ($\varepsilon_h$).

| Contacts type | $N_{nb}$ | $E_{nb}$, kcal/mol | $\varepsilon_h$, kcal/mol |
|---|---|---|---|
| intra-monomer contacts (α-tubulin) | 1340 | 2345 | 1.8 |
| intra-monomer contacts (β-tubulin) | 1240 | 2320 | 1.9 |
| intra-dimer contacts | 78 | 150 | 1.9 |
| longitudinal inter-dimer contacts | 38 | 37.6 | 1.0 |
| lateral inter-dimer contacts | 20 | 17.6 | 0.9 |

**Table S4: Major structural determinants of the α- and β-tubulin monomers involved in formation of inter-monomer contacts:** Analysis of contacts was performed for a 10 ns trajectory of equilibrium dynamics of the MT lattice (fragment MT8/13) in implicit solvent at $T = 300$ K. A pair of amino acids was assumed to form a binary contact if the distance between their $C_\alpha$-atoms was shorter than 8 Å. The secondary structure assignment is the same as in Refs. (15,16).

| Contacts type | Structure elements |
|---|---|
| intra-dimer contacts | α-tubulin: loops T3, T5, H6-H7, and H11-H12; helices H2, H6, and H11<br>β-tubulin: loops H1-S2, T4, T7, S8-H10, and H10-S9; helices H8 and H10;<br>β-strands S1 and S9 |
| longitudinal inter-dimer contacts | α-tubulin: loops T7 and H10-S9; helices H8, H10, and H12; β-strands S1 and S9 <br><br> β-tubulin: loops S2-H2, T3, T5, and H6-H7; helices H2 and H11; β-strand S3 |
| lateral inter-dimer contacts | 1st monomer (α- or β-tubulin): loops H1-S2, and T2; helix H3 <br><br> 2nd monomer (α- or β-tubulin): M-loop; helices H6, H9, and H10 |

**Supporting Movie Legends**

**Movie S1. Dynamic force spectroscopy *in silico*: Forced indentation of microtubule lattice between protofilaments.** The movie shows the forced indentation experiment *in silico* on the MT, in which a compressive force is applied in between the protofilaments at the indentation point 6 (side view). The MT cylinder (MT8/13) is positioned on a solid mica surface (not shown), as in Fig. 1B. The cantilever base (virtual particle) is moving with the velocity $v_f = 1.0$ µm/s perpendicular to the surface of the MT cylinder (not shown). As a result, the cantilever tip (gray colored sphere of radius $R_{tip} = 10$ nm) exerts pressure onto the outer surface of the MT, which undergoes a series of transformations: from continuous deformation at the early stage of indentation to the discrete transitions later in the sequence of events. These visible transitions include the dissociation of the lateral bonds between the protofilaments, and then longitudinal bonds between the tubulin dimers. The movie stops when the indentation depth reaches $X = 23$ nm. The duration of this indentation experiment is ~50 ms and the length of the movie is ~12 s (the movie is played ~240 times slower than experiment).

**Movie S2. Dynamic force spectroscopy *in silico*: Forced indentation and force-quenched tip retraction of microtubule lattice between protofilaments.** The movie shows initially the forced indentation of the MT (similar to Movie S1: MT8/13; indentation point 6, $v_f = 1.0$ µm/s, $R_{tip} = 10$ nm), in which a compressive force is applied in between the protofilaments (side view). After the indentation depth has reached $X = 20$ nm, the tip motion is reversed and the force is slowly quenched to zero. The movie stops after the cantilever tip reaches the initial point ($X = 0$ nm indentation). Partial MT cylinder restructuring is observed during the tip retraction, which includes re-formation of the lateral bonds but not longitudinal bonds. The duration of this indentation-retraction experiment is ~70 ms and the length of the movie is ~18 s (movie is played ~257 times slower than experiment).

**Movie S3. Dynamic force spectroscopy *in silico*: Force deformation of 16-dimers long microtubule protofilament.** The movie shows the forced deformation experiment *in silico*, in which a compressive force is applied to a single MT protofilament (PF16/1) clamped at both ends (side view). As in Movies S1 and S2, the cantilever base (virtual particle) is moving

perpendicular to the PF16/1surface with velocity $v_f = 0.2$ µm/s (not shown). This results in the cantilever tip (gray colored sphere of radius $R_{tip} = 10$ nm) exerting pressure on the protofilament. The movie shows that the pressure increase leads to the protofilament bending (initially) and to the dissociation of longitudinal bonds (later). The duration of this deformation experiment is ~90 ms and the length of the movie is ~15 s (movie is played ~167 times slower than experiment).

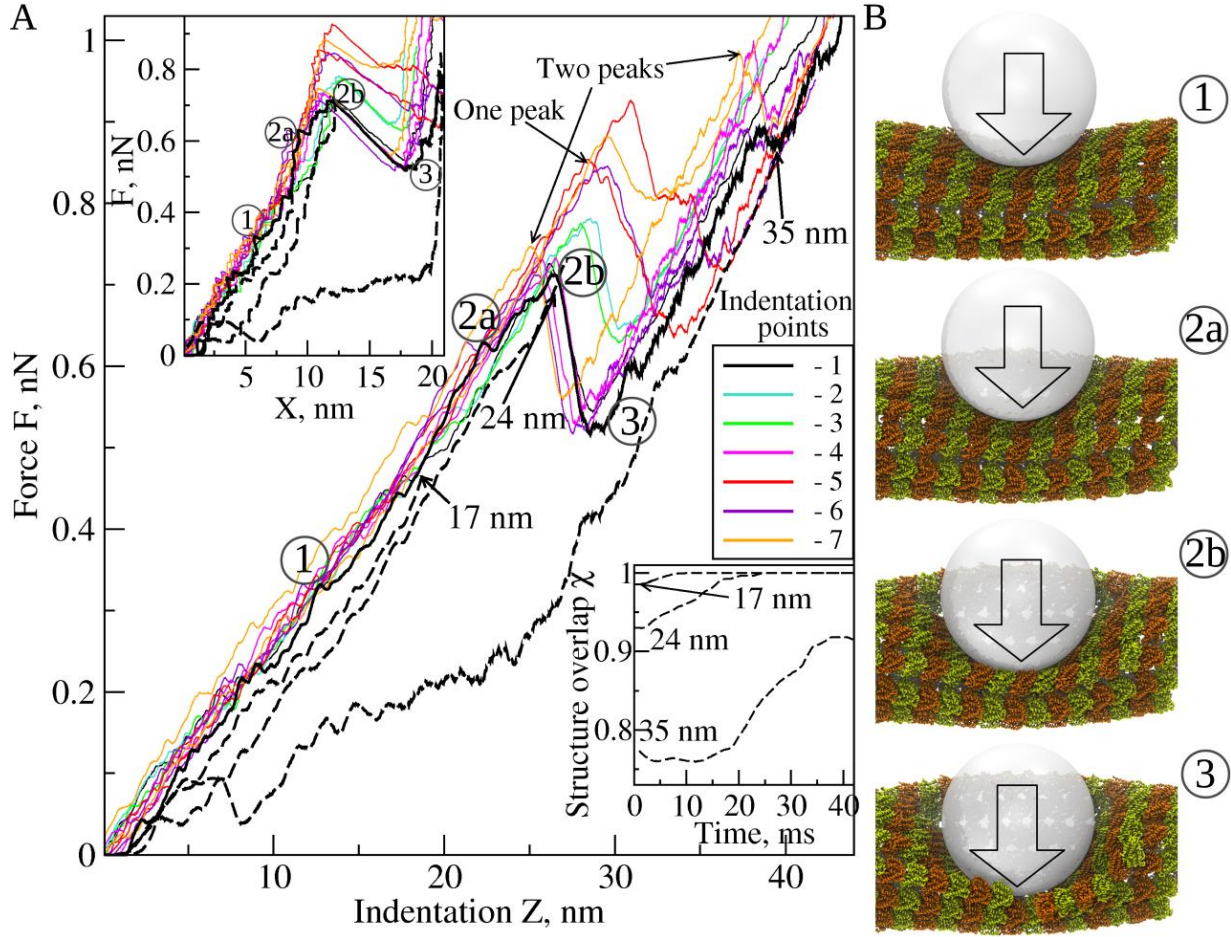

**Fig. S1. Forced indentation of MT lattice as in Fig. 2 but for a larger tip. A:** The force-deformation (*FZ*) curves for 7 indentation points (Fig. 1C) obtained with $v_f = 1.0$ μm/s and $R_{tip} = 15$ nm (see Fig. 2 for results obtained with smaller tip of $R_{tip} = 10$ nm). The curves are shown in different color for clarity. Dashed black curves represent the *FZ* profiles for the tip retraction simulations, which followed the forward indentations (solid black curve) with $Z = 17$, 24, and 35 nm as initial conditions. *The top inset* shows the corresponding *FX* curves for the forward indentation (solid curves; colors as in A) and backward tip retraction (dashed black curves). *The bottom inset* shows the time profiles of the structure overlap $\chi$ for MT lattice restructuring during tip retraction (starting from $Z = 17$, 24, and 35 nm indentation). **B:** The MT structures 1, 2a, 2b, and 3, illustrating the mechanism of MT deformation and collapse (as in Fig. 2 in the main text), which correspond to the accordingly numbered regions in the force-deformation curves in **A.**

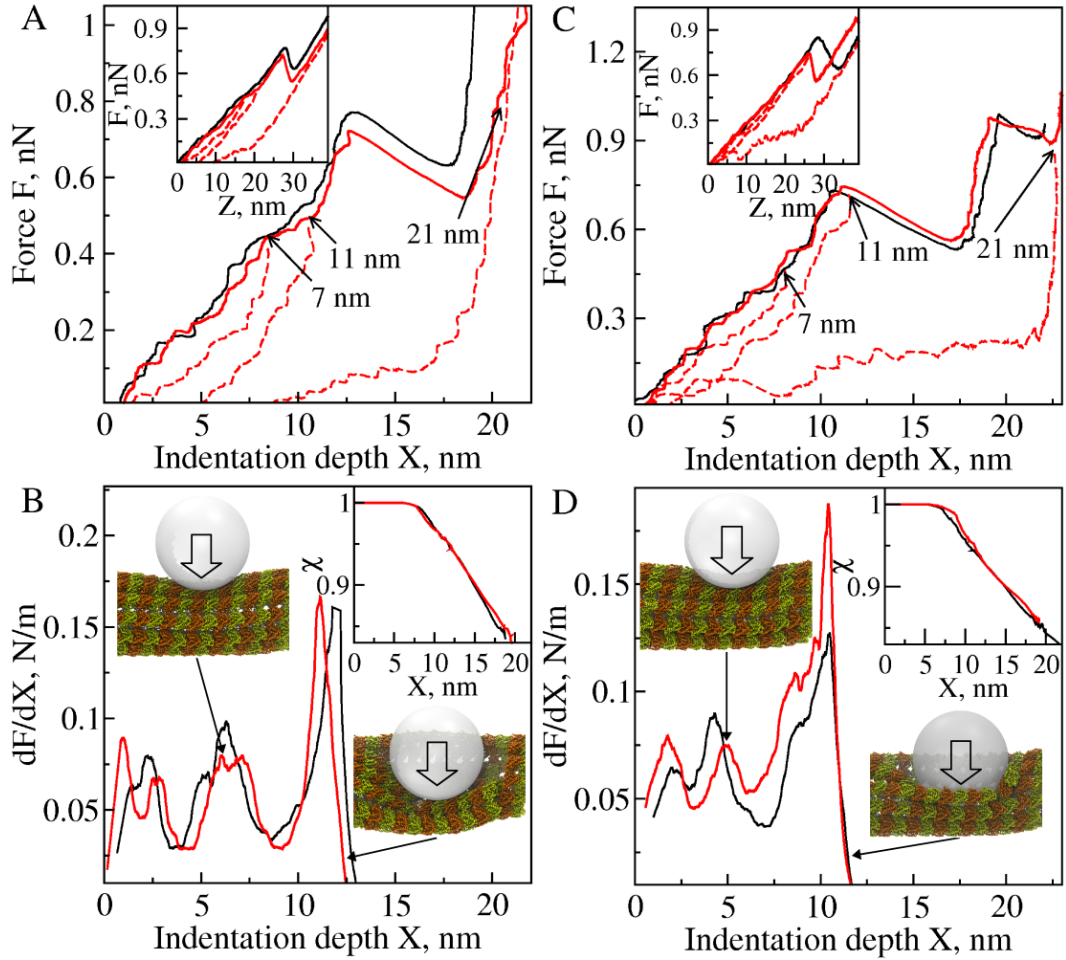

**Fig. S2. Force spectra for indentation and retraction of the MT lattice *in silico* and corresponding structure alterations.** Representative results for indentation points 3 (black) and 1 (red) in **A** and **B**, respectively, and for indentation points 5 (black) and 7 (red) in **C** and **D**, respectively, obtained with $v_f = 1.0$ μm/s and $R_{tip} = 15$ nm (see Fig. 3 for results obtained with smaller tip of $R_{tip} = 10$ nm). **A** and **C:** The *FX* curves (*FZ* curves are in the *inset*) for forward indentation (solid black and red curves). Curves for backward tip retraction (dashed red lines) were generated using the structures obtained from forward indentation for $X = 7$, 11, and 21 nm (indicated on the graphs). **B** and **D:** The slope of the *FX* curves $dF/dX$ for the force spectra from panels **A** and **C**. Snapshots show the side-views of the MT structure before dissociation of the lateral bonds and after dissociation of the longitudinal bonds. *The insets* show the profiles of structure overlap $\chi$ vs. $X$, which demonstrate that the MT lattice in the fully collapsed state ($X > 20$ nm) is ~80-85% similar to the native state.

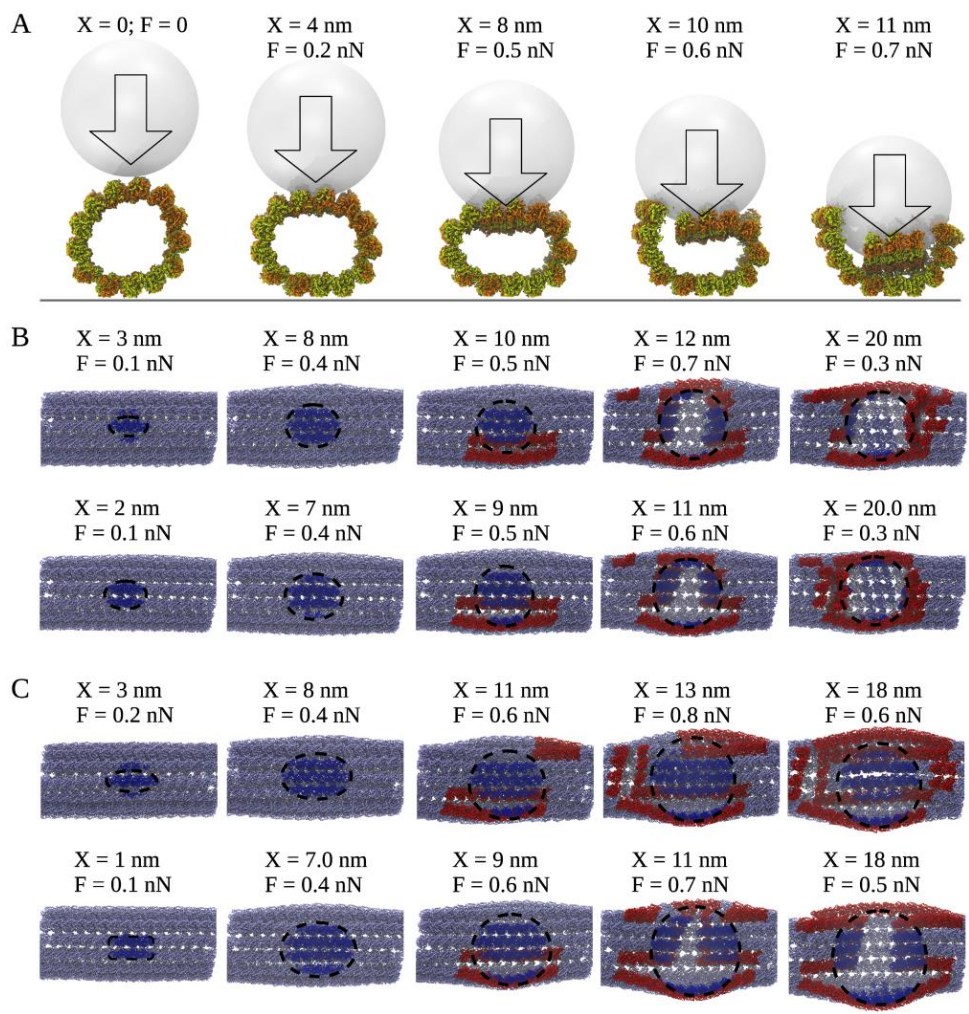

**Fig. S3: Dynamic structure changes observed during mechanical compression of MT fragment MT8/13**. **A:** Profiles of the MT lattice viewed along the MT cylinder axis for different extent of indentation obtained with $R_{tip}$ = 15 nm (indentation point 7; see Fig. 1C). **B** and **C** show top views of the MT lattice for indentation points 3 (upper raw in **B**) and 6 (lower raw in **B**) obtained with smaller tip ($R_{tip}$ = 10 nm), and for the indentation points 3 (upper raw in **C**) and 4 (lower raw in **C**) obtained with larger tip ($R_{tip}$ = 15 nm). During compression, the MT structure (light blue) becomes deformed and there is an associated increase in the MT-tip contact area (encircled dark blue area). Subsequent increase in the compressive force results in the dissociation of lateral and longitudinal bonds (tubulin monomers with disrupted lateral and/or longitudinal interface(s) are shown in dark red).

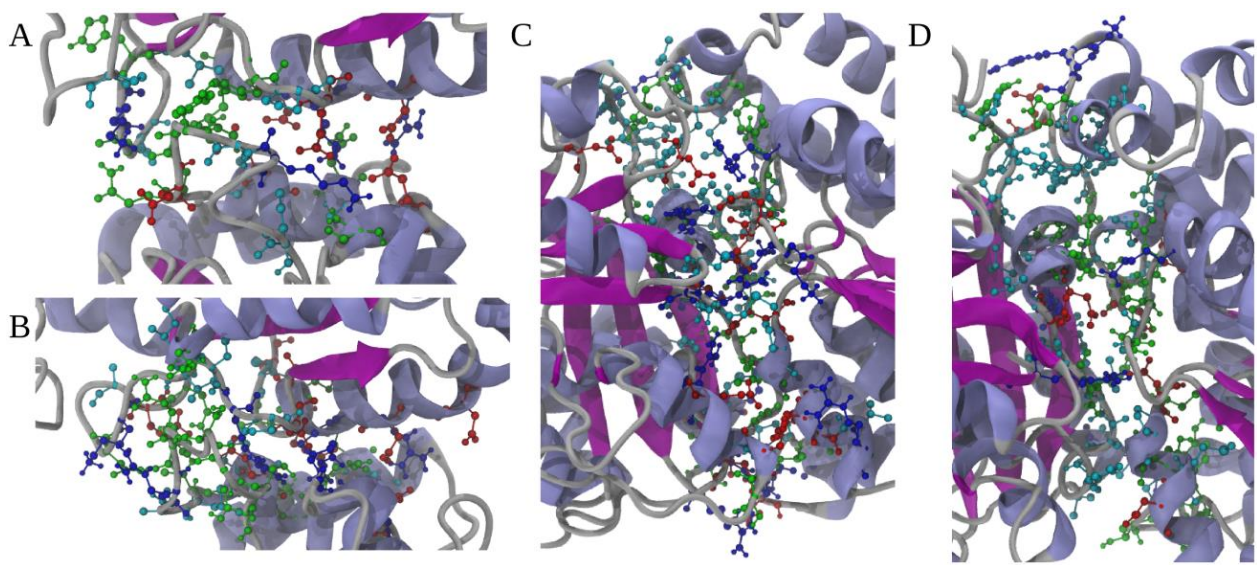

**Fig. S4: Atomic representation of the interfaces, corresponding to lateral α-α (A) and β-β tubulin bonds (B), and the longitudinal intra-dimer (C) and inter-dimer bonds (D)**. The most stable side-chain contacts, i.e. the contacts that persist for 5 ns of simulations at equilibrium, are shown using the "balls-and-sticks" representation. The binary contact is defined as the contact between amino acids for which the distance between their $C_\alpha$-atoms does not exceed 8 Å (see Materials and Methods). Color denotation: acidic side chains are in red, basic side-chains in blue, polar residues in green, and hydrophobic residues are in cyan. Color denotation for the secondary structure: α-helices (blue), β-sheets (red), and loops/turns (gray). The lateral interface between the α-tubulins contains ~19 side-chain contacts (**A**); the lateral interface between the β-tubulins contains ~21 contacts (**B**); the longitudinal intra-dimer interface is stabilized by ~78 contacts (**C**); and the longitudinal inter-dimer interface is stabilized by ~38 contacts (**D**).

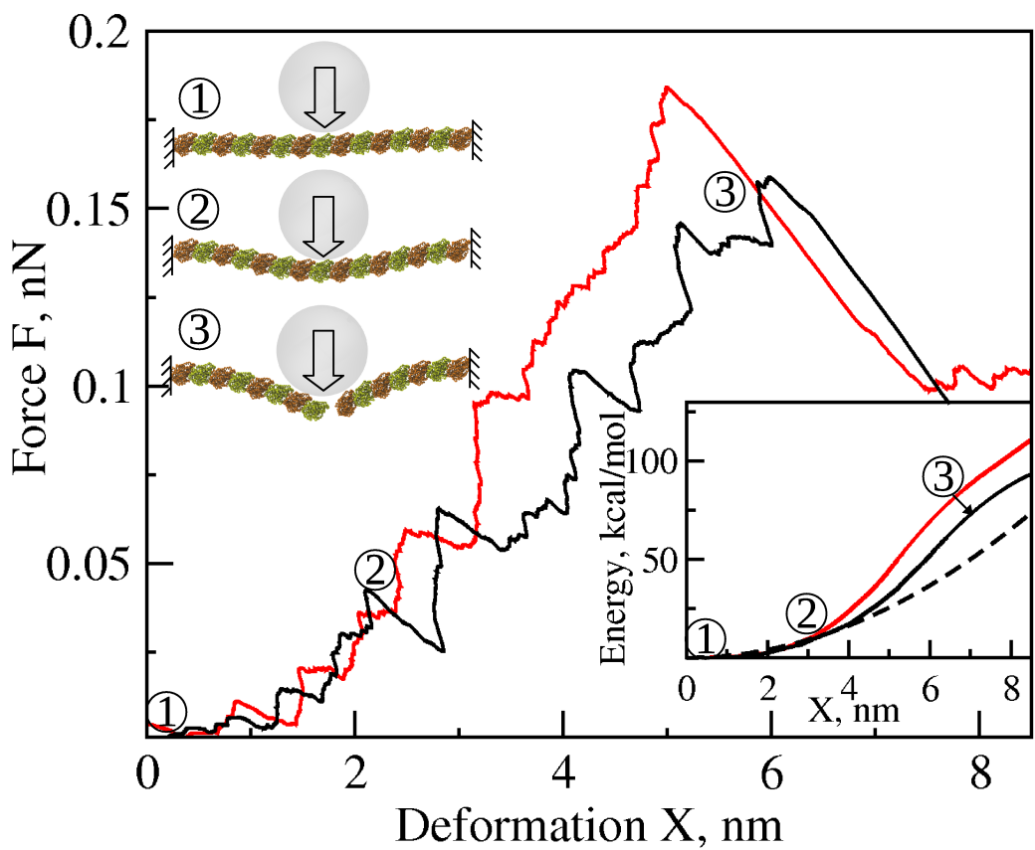

**Fig. S5: Force-deformation spectra for single protofilament fragment of the MT cylinder.** Shown are the typical examples of the force spectra for protofilament fragment PF8/1 (black and red solid curves) obtained by using the cantilever velocity $v_f = 0.2$ μm/s and spherical tip radius $R_{tip} = 10$ nm. Structures numbered 1-3, which show deformation progress, correspond to the black *FX* curve and represent the native state, the weakly bent state ($X = 2$ nm), and the strongly bent dissociated state ($X = 6$ nm), respectively. The tip shown with the vertical arrow deforms the protofilament until the dissociation of the longitudinal bond occurs. *The inset* shows the corresponding profiles of bending energy as a function of deformation for the estimation of the flexural rigidity. The dashed black curve is a fit of the quadratic function $V \sim X^2$ to the black curve of the energy *V* as a function of *X*, which shows the validity of the harmonic approximation used to calculate *EI* for protofilament fragments PF8/1, PF16/1, PF24/1, and PF32/1 (Table 3 in the main text).

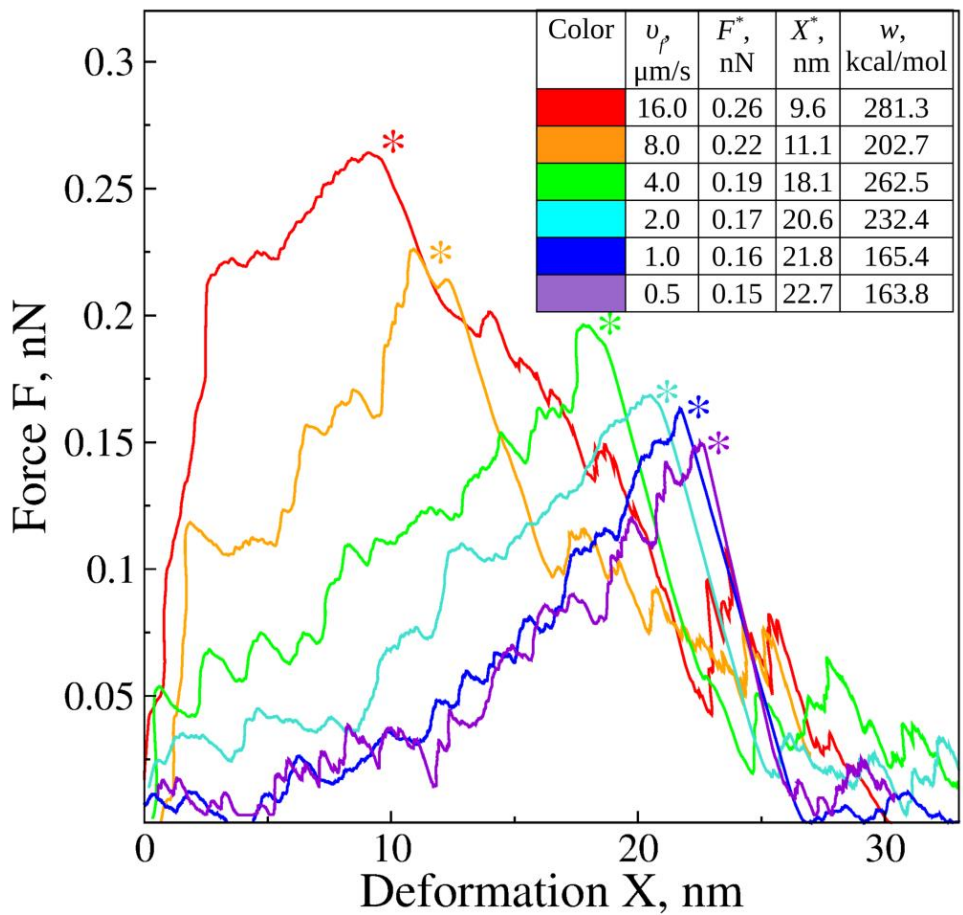

**Fig. S6: Dependence of force-deformation spectra for single protofilament fragment on cantilever velocity.** Shown in different colors for clarity are the typical examples of the force spectra (*FX* curves) for 24-dimer long protofilament fragment PF24/1 obtained by using the cantilever velocity $v_f$ = 0.5 (violet), 1.0 (blue), 2.0 (turquoise), 4.0 (green), 8.0 (orange), and 16.0 μm/s (red color) and spherical tip ($R_{tip}$ = 5 nm). The asterisks above the *FX* curves mark the onset of dissociation of the protofilament fragment, which occurs at the critical force $F^*$ and critical deformation $X^*$. *The inset* is the summary of the values of $F^*$, $X^*$, and critical deformation work $w^* = \int_0^{X^*} F(X)\,dX$ (area under the *FX* curve up to $X = X^*$) obtained for different $v_f$. As $v_f$ decreases, the *FX* curves change less and less. We also observe the convergence of $F^*$, $X^*$, and $w^*$ with the decreasing velocity $v_f$, each approaching some stable values. This shows that *in silico* indentation experiments carried out at $v_f$ = 0.5-1.0 μm/s cantilever velocity correspond to near-equilibrium conditions of the compressive force application.

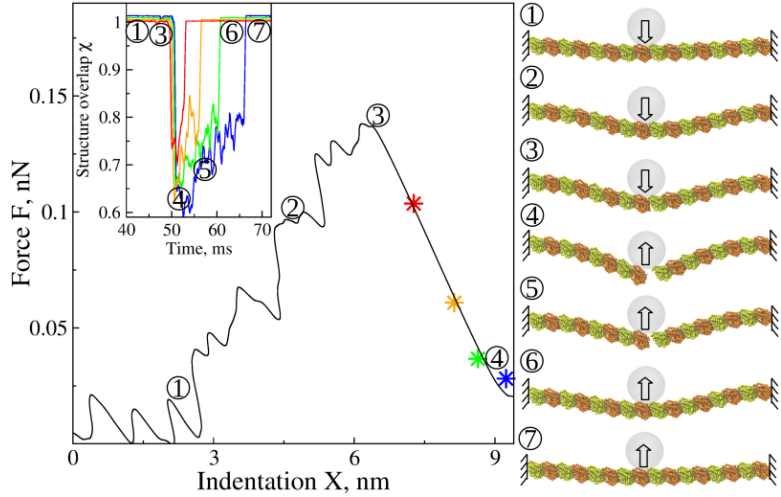

**Fig. S7: Reversibility of longitudinal bonds' dissociation as observed for single protofilament fragment of the MT cylinder.** Shown is the force-deformation spectrum for protofilament fragment PF8/1 (*FX* curve) obtained with $v_f$ = 0.5 µm/s and $R_{tip}$ = 5 nm. Structures numbered 1-4 which show deformation progress represent the near-native state ($X$ = 2.0 nm; structure 1), weakly bent state ($X$ = 4.5 nm; structure 2), strongly bent un-dissociated state ($X$ = 6.5 nm; structure 3), and strongly bent dissociated state ($X$ = 8.5 nm; structure 4), respectively. The tip (with vertical arrow) deforms the protofilament until the longitudinal bond dissociates. The asterisks mark the points for $X$ = 7.0 nm (red), 8.0 nm (orange), 8.5 nm (green), and 9.2 nm deformations (blue), which correspond to the conformations of the strongly bent dissociated state of PF8/1. These 4 conformations were further used as initial structures in the simulations of force-quench retraction. *The inset* shows the time-profile of the structure overlap function $\chi$ (see Data analysis section) for the 4-th and 5-th dimer of PF8/1, which captures dissociation and subsequent re-formation of the longitudinal bond between the 4-th and 5-th dimer for all initial conditions used. Beyond the 50 ms time point, at which the bond dissociates (sharp drop in $\chi$), the bond re-forms and the protofilament unbends over the 3, 6, 11, and 16 ms timeframe ($\chi$ approaches the unity) for all bond-dissociated structures corresponding to $X$ = 7.0, 8.0, 8.5, and 9.2 nm deformations, respectively. Snapshots numbered 5-7, which correspond to $X$ = 8.5 nm deformation (green asterisk), show the dynamics of bond re-formation and protofilament re-structuring observed in tip-retraction simulations. Hence, the longitudinal bond dissociation in short protofilament fragments (such as PF8/1) is reversible on the 10-20 ms timescale of our computational experiments.